\documentclass[fleqn,usenatbib]{mnras}

\usepackage{newtxtext,newtxmath}

\usepackage[T1]{fontenc}
\usepackage{ae,aecompl}

\usepackage{graphicx}	
\usepackage{amsmath}	
\usepackage{multirow}

\makeatletter
\patchcmd{\NAT@test}{\else \NAT@nm}{\else \NAT@nmfmt{\NAT@nm}}{}{}

\DeclareRobustCommand\citepos
  {\begingroup
   \let\NAT@nmfmt\NAT@posfmt
   \NAT@swafalse\let\NAT@ctype\z@\NAT@partrue
   \@ifstar{\NAT@fulltrue\NAT@citetp}{\NAT@fullfalse\NAT@citetp}}

\let\NAT@orig@nmfmt\NAT@nmfmt
\def\NAT@posfmt#1{\NAT@orig@nmfmt{#1's}}
\makeatother

\newcommand{\Msun}{\mathrm{M}_\odot}
\newcommand{\nvar}{n_{\rm var}}
\newcommand{\nbr}{n_{\rm br}}
\newcommand{\nsnaps}{n_{\rm snap}}
\newcommand{\Mvir}{\mathrm{M_{vir}}}

\title[Merger tree construction using deep learning]{A deep learning approach to halo merger tree construction}

\author[S. Robles et al.]{
Sandra Robles,$^{1,2,3}$\thanks{E-mail: sandra.robles@kcl.ac.uk (SR)}
Jonathan S. G\'omez,$^{1, 4}$\thanks{E-mail: jsgomez1@uc.cl (JSG)}
Ad\'in Ram\'irez~Rivera,$^{5}$ 
 Nelson D. Padilla$^{6}$ and
\newauthor \ Diego Dujovne$^{7}$
\\
$^{1}$Departamento de F\'isica Te\'orica, Universidad Aut\'onoma de Madrid, E-28049 Cantoblanco, Madrid, Spain.\\
$^{2}$Theoretical Particle Physics and Cosmology Group, Department of Physics, King’s College London, Strand, London, WC2R 2LS, UK\\
$^{3}$ARC Centre of Excellence for Dark Matter Particle Physics, 
School of Physics, The University of Melbourne, Victoria 3010, Australia.\\
$^{4}$Instituto de Astrof\'isica, Pontificia Universidad Cat\'olica de Chile, Av. Vicu\~na Mackenna 4860, Santiago, Chile.\\
$^{5}$Department of Informatics, University of Oslo, Gaustadall\'een 23 B, N-0373, Oslo, Norway.\\
$^6$Instituto de Astronom\'ia Te\'orica y Experimental, UNC-CONICET, C\'ordoba, X5000BGR, Argentina\\
$^{7}$Escuela de Inform\'atica y Telecomunicaciones, Universidad Diego Portales, Av. Ej\'ercito 441, Santiago, Chile.
}

\date{Accepted 2022 June 1. Received 2022 May 31; in original form 2022 April 13}

\pubyear{2022}

\begin{document}
\label{firstpage}
\pagerange{\pageref{firstpage}--\pageref{lastpage}}
\maketitle

\begin{abstract}
A key ingredient for semi-analytic models (SAMs) of galaxy formation is the mass assembly history of haloes, encoded in a tree structure. The most commonly used method to construct halo merger histories is based on the outcomes of high-resolution, computationally intensive N-body simulations. We show that machine learning (ML) techniques, in particular Generative Adversarial Networks (GANs), are a promising new tool to tackle this problem with a modest computational cost and retaining the best features of merger trees from simulations. 
We train our GAN model with a limited sample of merger trees from the Evolution and Assembly of GaLaxies and their Environments  (EAGLE) simulation suite, constructed using two halo finders--tree builder algorithms:  SUBFIND--D-TREES and ROCKSTAR--{\sc ConsistentTrees}.
Our GAN model successfully learns to generate well-constructed  merger tree structures with high temporal resolution, and to reproduce the statistical features of the sample of merger trees used for training,  when considering up to three variables in the training process. These inputs, whose representations are also learned by our GAN model, are mass of the halo progenitors and the final descendant, progenitor type (main halo or satellite) and distance of a progenitor to that in the main branch. 
The inclusion of the latter two inputs greatly improves the final learned representation of the halo mass growth history, especially for SUBFIND-like ML trees. When comparing equally sized samples of ML  merger trees with those of the EAGLE simulation, we find better agreement for SUBFIND-like ML  trees. 
Finally, our GAN-based framework can be utilised to construct merger histories  of low- and intermediate-mass haloes, the most abundant in cosmological simulations. 
\end{abstract}




\section{Introduction}

The dark energy dominated dark matter model,  $\Lambda$ cold dark matter ($\Lambda$CDM), provides a successful theoretical framework for understanding and simulating galaxies,  which are believed to inhabit dark matter (DM) haloes. Galaxy formation and evolution are  complex non-linear problems, theoretically and  numerically. Gas-dynamical and radiative processes, such as the formation of stars and black holes as well as their respective feedback, have to be taken into account for simulated galaxies to resemble reality. Over the past decades,  two different strategies have been developed to tackle this problem:  hydrodynamical simulations \citep{Carlberg1990,Katz1992} and semi-analytic models (SAMs) \citep{Cole1991,White1991,Croton2016}.

Hydrodynamical simulations directly address a wide range of dynamical scales and solve numerically the combined non-linear N-body and hydrodynamic equations describing the formation of galaxies. They provide many advantages, among them, self-consistent evolution of DM and baryonic components, high resolution of the latter component, and, furthermore, they simultaneously model galaxies and the intergalactic medium. Even if high resolution simulations are massively parallelised and run on supercomputers, they are extremely computational resource intensive. Alternatives to circumvent this limitation are semi-analytic modelling and generating DM halo catalogues with less orders in Lagrangian perturbation theory \citep{Munari2017}. These models are established tools for connecting the predicted hierarchical growth of DM haloes, the halo merger trees, to the observed properties of the galaxy population \citep[see e.g.][]{Cole2000, Somerville2008, Guo2011}. It is worth noting that even though both hydrodynamical simulations and semi-analytic models rely on subgrid physical models to account for processes that cannot be directly simulated, they are more approximate in the latter models. 

In the standard model of cosmology, galaxies form in DM haloes collapsed from tiny overdensities. Large DM haloes are formed by the collapse and merger of smaller structures or progenitors. Thus, galaxy formation and evolution are driven by the halo merger history. If the progenitors contain galaxies, halo mergers eventually give rise to galaxy mergers. The merging and hierarchical formation history of DM haloes can be obtained in cosmological simulations by tracing all halo progenitors and storing them in tree structures, commonly referred to as `halo merger trees.'

There is another method to produce halo merger trees, which is based on the extended Press-Schechter formalism \citep{Bond1991} and Monte Carlo simulations \citep{Kauffmann19931, Kauffmann19932, Cole1994}. This is a  simple but relatively effective framework for the description of the mass history of particles in a hierarchical Universe. Its main advantage is a rapid  merger tree construction in large volumes with high mass resolution  \citep{Lacey1993,Somerville1999,Cole2000,Somerville2008,Benson2010,Ricciardelli2010}. However, only one tree can be  built at a time and often this framework yields  results that are  in disagreement with simulations \citep{Jiang2014J}.

Cosmological N-body simulations are a powerful and well established tool for studying theories of cosmic structure formation and for making predictions that can be compared directly to observations. Despite being computationally intensive,  high-resolution DM only (N-body) simulations yield a more realistic evolutionary history of the haloes and are capable of producing thousands of merger trees at once  \citep{Roukema1997,Kauffmann1999,Okamoto2001,Hatton2003,DeLucia2004,Croton2006,Bower2006,Guo2011}. This method, however, is not exempt of subtleties.  The most relevant being the  dependence on the clustering algorithms  employed to find haloes and to build trees \citep{Knebe2011,Avila2014,Gomez2021}. Another important caveat is the  mass resolution limit, poorly resolved haloes or dense environments can be problematic for identifying substructures for some halo finders (or substructure finders; \citealt{Muldrew2011, Onions2013, Elahi2013}). Contrary to  haloes with a small number of particles, the assembly history of massive haloes can be traced back to progenitors whose masses  are a small fraction of the mass of the final descendant.

Despite the difficulties of studying the evolution of galaxies, merger trees of DM haloes play an important role in modern galaxy formation theory. They are the backbone of SAMs. SAMs populate dark matter haloes in cosmological simulations by using analytical approximations to self-consistently model the evolution of galaxies throughout cosmic time. Due to the flexibility of SAMs to explore physical phenomena, they are best suited to compare theoretical predictions with galaxy surveys \citep{Lagos2018}. However, the necessary condition for a SAM to produce galaxy formation and merger histories is  to have a complete sample of well-constructed and realistic merger trees.

In recent years, deep learning algorithms, a subset of machine learning (ML) techniques that compose functions from parameterized operations that enforce few inductive biases over the parameterization, have increasingly been used in astrophysics  because they  can process large sets of data and extract features from them by observing patterns in the data. 
In particular, convolutional neural networks (CNNs), a deep learning model designed to learn spatial hierarchies of features, have been employed, among others, to measure the dynamical mass of galaxy clusters \citep{Ho2019},  map N-body and 
hydrodynamical  simulations \citep{Wadekar:2021}  and thereby exploring the halo-galaxy connection, to morphologically classify galaxies 
\citep{Dieleman:2015,Kim:2017, Barchi2020, Cavanagh2021}, and  segment and classify the large-scale structure of the Universe  \citep{AragonCalvo:2019}.

With the aim of providing  a new framework for halo merger tree construction, taking advantage of the best features of large-volume simulations, but with a modest computational expense,  we extend the deep CNN model designed and tested by \citet{Robles:2019nfk}. This model is based on a Generative Adversarial Network (GAN) \citep{Goodfellow2014}, which we  train  with merger trees from the largest dark matter only simulation in the Evolution and Assembly of GaLaxies and their Environments (EAGLE) simulation suite \citep{Shaye2015,Crain2015}. 
GANs are an unsupervised deep learning technique, characterised by training a pair of neural networks in competition with each other in a sort of minimax game.
Finding new applications for GANs is currently an active area of research. They have been employed mainly for computer vision applications such as image generation, editing and classification. Recently, they have been proved to be useful to solve computational and data intensive tasks in physics. We propose here a new application in astrophysics  that can help SAMs of galaxy formation to more reliably simulate large upcoming galaxy surveys and allow us to more rapidly compare theory with observations.

We perform a series of experiments including  up to three input variables in the training process. This choice of variables that describe halo merger trees is motivated by SAMs~\citep{Cole2000,Benson2012,Croton2016,Cora2018}. Namely these quantities are 
mass growth, distance between merging progenitors and the progenitor type: main or satellite halo.   These experiments were designed to test the capabilities of our GAN model and the relevance of each input to the well reconstruction of the mass growth history of haloes. To validate the correct construction of the ML-generated  trees, we compare statistically significant samples of ML-generated with `real' trees, where `real' stands for merger trees extracted from the EAGLE simulation and ML-generated trees are  the outputs of our GAN model, and find very good agreement. 

This paper is organised as follows. In Section~\ref{sec:simumethods}, we briefly give details of the EAGLE  simulation, halo finder and tree builder algorithms utilised to extract merger trees from simulations. In Section~\ref{sec:GAN}, we outline the GAN model employed to generate merger trees and the statistical measures used to assess the quality of these trees. Details of the GAN architecture can be found in Appendix~\ref{sec:GANarchitecture} and additional examples of ML generated trees in Appendix~\ref{sec:examples}. Our results are presented in Section~\ref{sec:results}. Concluding remarks are given in Section~\ref{sec:conclusions}.

\section{Merger Trees from Simulations}
\label{sec:simumethods}

\begin{table}
    \small\addtolength{\tabcolsep}{-5pt}
    \centering
    \caption{Parameters and technical specifications of the N-body simulations used in this work.}    
    \begin{tabular}{lccc}
    \hline
    Parameter        & \multicolumn{2}{c|}{Physical meaning}                   & Value   \\
    \hline
    $\Omega_m$       & \multicolumn{2}{c|}{Present fractional matter density}  & 0.307   \\
    $\Omega_\Lambda$ & \multicolumn{2}{c|}{Present fractional vacuum energy density }                       & 0.693   \\
    $h$              & \multicolumn{2}{c|}{$H_0/(100$ km s$^{-1}$ Mpc$^{-1})$} & 0.6777  \\
    $n_s$            & \multicolumn{2}{c|}{Primordial power spectral index}    & 0.9611  \\
    $\sigma_8$       & \multicolumn{2}{c|}{rms linear density fluctuation}     & 0.8288  \\
    \hline
    Simulation    &	$L_{\mathrm{box}}$ (cMpc) & $N_{\mathrm{p}}$ & $m_{\mathrm{dmp}}$($h^{-1}\Msun$)\\
    \hline  
    EAGLE100      &  100                     & $1504^3$         & $6.57 \times 10^6$                      \\
    \hline
    \end{tabular}
    \label{tab:cosmo_params}
\end{table}

In order to generate halo merger trees  of dark matter haloes using neural networks, a training dataset of `real' merger trees is required, which we select from the  EAGLE simulation suite. This suite consists of several  hydrodynamical simulations \citep{Shaye2015}, but also contains dark matter only (DMO) versions of the reference simulations. The DMO simulations were obtained using the same initial conditions and the same resolutions as the reference models, DMO simulations and their hydrodinamical counterparts  were later matched \citep{Schaller2015}. We use the largest DMO simulation, which has  
 a cubic periodic volume  of 100 co-moving Mpc side (hereafter referred to as E100)  and with a dark matter particle (DMP) mass $m_{\mathrm{dmp}}=6.57 \times 10^6 h^{-1} \Msun$. The simulation adopts the \cite{planck2014} cosmological parameters, listed in Table~\ref{tab:cosmo_params}. E100 has a high temporal resolution of 201 snapshots, numbered from 0 to 200, distributed between $z=20$ and $z=0$, respectively. This simulation supplies a large enough amount of merger trees for training purposes and  its high temporal resolution helps us to better distinguish  pre-merger phases.

To be able to compare different learning processes of our neural network architecture, we shall use two distinct halo merger tree databases, obtained from the aforementioned simulation by applying two clustering algorithms to construct merger trees in the E100 simulation. This process is performed in two steps: 
\begin{enumerate}
    \item First, the halo finder identifies all the haloes in each snapshot using the dark matter particles of the simulation.
    \item Next, the tree builder constructs links between haloes across  different snapshots.
\end{enumerate}
This is the general method  to build halo merger trees, called by convention, halo finder--tree builder.

We employ halo merger trees identified with two combinations of halo finder--tree builder listed in Table~\ref{table: halo finders/tree builders}. As a first  step, both halo finders make use of the Friends-of-Friends standard algorithm (hereinafter FoF). Then, in subsequent steps they perform a more refined search of haloes and their substructures, using more sophisticated techniques.  
For our study, we use the SUBFIND \citep{Springel2001} halo finder that identify 3D overdensities (i.e., they consider only the position of the particles) and ROCKSTAR \citep{Behroozi2013a} halo finder that identifies 6D overdensities (basically phase-space, i.e., particle positions and velocities).

It is worth mentioning that there exist many other halo finder and tree builder codes in the literature. Both algorithms are at least equally important  to produce  well constructed  merger trees~\citep{Avila2014}. Yet they are both subject to issues which have been studied in detail. \citet{Knebe2011,Knebe2013} compared different halo finders, and investigated the sources of errors and discrepancies among them when identifying haloes and their substructure. Several methods to build merger trees were compared using the same halo catalogue by \citet{Srisawat2013}. 
They concluded that a reliable tree builder should be able to trace particle transfers in order to match haloes between adjacent snapshots, and skip at least one snapshot to  correct for missing haloes.
They also found that different tree builders even using the same halo catalogue yield distinct halo growth histories.  This could alter galaxy properties when utilising them in SAMs \citep{Lee2014}.  The effect of mass and temporal resolution on merger tree construction has also been investigated  \citep{Wang2016}. An important recommendation from this study is  that when merger trees are built using  more than 100 snapshots, which is precisely our case, the tree builder algorithm should at least be able to deal with issues in the halo catalogue.  
In the following subsections, we briefly provide details of each of the two halo finder--tree builders and the terminology adopted in this work.

    \begin{table}
    \centering
    \caption{Halo merger tree builders considered in this work. The first column gives the halo finder algorithm and the second column the corresponding tree builder.}
    \begin{tabular}{cc}
    \hline
    Halo finder           &  Tree builder         \\
    \hline
    SUBFIND               & D-TREES               \\
    \citep{Springel2001}  & \citep{Jiang2014,Qu:2017}     \\
    \hline
    ROCKSTAR              & {\sc ConsistentTrees}       \\
    \citep{Behroozi2013a} & \citep{Behroozi2013b} \\
    \hline
    \end{tabular}
    \label{table: halo finders/tree builders}
    \end{table}

\subsection{SUBFIND -- D-TREES}
SUBFIND \citep{Springel2001} is a self-bound particle substructure finder which first identifies particle groups using the Friends-of-Friends algorithm with linking length $b = 0.2$ times the mean inter-particle separation. In order to identify the gravitationally bound haloes from each group, a local density is estimated for each particle with an adaptive kernel interpolation that uses a prescribed number of smoothing neighbours. Starting from isolated density peaks,  particles are added in sequence of decreasing density. Whenever a saddle point in the global density field is reached, such that it connects two disjoint overdense regions, the smaller candidate is treated as a separate satellite halo. All substructure candidates are subjected to an iterative unbinding procedure with a tree-based calculation of the potential.

The D-TREES algorithm \citep{Jiang2014} is a tree builder created to be used together with SUBFIND, that finds merger histories taking into account the uncertainty in the definition of a halo and possible loss of particles. D-TREES first considers two consecutive snapshots in the simulation. Each halo is  identified as the progenitor of whichever halo at the next snapshot contains the largest fraction of its particles. This process is repeated for all pairs of consecutive snapshots. It is then straightforward to trace the merger history of each halo that exists at the final output time. Then, the algorithm checks that the most bound particle of a halo remains a member of the descendant halo, and also requires that the majority of the constituent particles of a halo are present in its descendant at the next output time. If this is not satisfied, D-TREES chooses the most bound particle from those that are in a halo at the later output time that contains {the largest number} of the progenitor particles.
Specifically, the merger trees used in this work have been obtained using the adaptation of the  D-TREES algorithm to the EAGLE simulation~\citep{Qu:2017}. 

\subsection{ROCKSTAR -- {\sc ConsistentTrees}}
ROCKSTAR, Robust Overdensity Calculation using K-Space Topologically Adaptive Refinement, \citep{Behroozi2013a}, is a phase-space halo finder\footnote{\href{https://bitbucket.org/gfcstanford/rockstar}{https://bitbucket.org/gfcstanford/rockstar}} designed to maximise halo consistency across time-steps. The algorithm first selects particle groups with a 3D Friends-of-Friends variant with a very large linking length ($b = 0.28$). For each FoF group, particle positions and velocities are divided (normalised) by the group position and velocity dispersion, giving a natural phase-space metric. Then, for each main group, ROCKSTAR builds a hierarchy of FoF subgroups. Thus, the metric ensures an adaptive selection of overdensities at each successive level of the FoF hierarchy. This  process is repeated for each subgroup; i.e. renormalisation, a new linking-length, and a new level of substructure are calculated for subgroups. When this is completed, ROCKSTAR converts FoF subgroups into seed haloes beginning at the deepest level of the hierarchy. If a particular group has multiple subgroups, then particles are assigned to subgroup seed haloes based on their phase-space proximity. This process is repeated at all levels of the hierarchy until all particles in the base FoF group have been assigned to haloes.

The {\sc ConsistentTrees} algorithm\footnote{\href{https://bitbucket.org/pbehroozi/consistent-trees}{https://bitbucket.org/pbehroozi/consistent-trees}} \citep{Behroozi2013b}, part of the ROCKSTAR package, first matches haloes between snapshots by identifying descendant haloes as those that have the maximum number of particles from a given progenitor. It then attempts to clean up this initial guess, either by correcting erroneous links or by adding missing haloes. To this end,  {\sc ConsistentTrees} simulates the gravitational motion of the set of haloes according to their known positions, velocities and mass profiles as returned by the halo finder (ROCKSTAR). Then, for  haloes in any given simulation snapshot, expected positions and velocities at an earlier snapshot can be inferred. 
To correct for missing haloes, {\sc ConsistentTrees} utilises halo trajectories from gravitationally evolving positions and velocities of haloes across time-steps. Thus, by using kinematic information from surrounding snapshots, it can correct for missing or extraneous haloes. 
This is required when ROCKSTAR can no longer find a satellite because it is too close to the host's centre, but the satellite is not fully merged so that a correction is needed to account for the missing satellite. In this way, {\sc ConsistentTrees}  modifies the original ROCKSTAR halo catalogue by adding missing haloes and their corresponding  links.

\subsection{Terminology}
In this subsection, we briefly define the terminology used in the following sections. A halo is a gravitationally bound structure as returned by the halo finder. Haloes can contain self-bound substructures or subhaloes. In that case, the halo that contains the particle with the lowest value of the gravitational potential is the main halo, and the remaining haloes are satellites. 

A merger tree, see Figs.~\ref{fig:mergertree} and \ref{fig:mergertreestruct},  is a graph of chronologically ordered set of haloes, 
the outputs of the tree builder. This tree represents the mass growth of a halo (the final descendant) over time (snapshots). The final descendant is located at the top node of the tree ($z=0$). The remaining haloes in a merger tree are dubbed   progenitors. A distinctive feature of a merger tree is the so-called main branch,\footnote{Also called trunk.} the largest branch of the tree that  contains the most massive progenitor, which is expected to be the main halo. 
Other branches emerge from the main branch connecting progenitors, which can be main or satellite haloes,  backwards in time. When two or more branches fuse to become one, a merger has occurred.  

\section{Halo merger tree generation}
\label{sec:GAN}

In this section, we outline the halo merger tree generation process implemented here,  which uses a deep-learning-based generative model.  We also define  statistical measures that help us to validate the appropriate construction of the ML generated trees. 

\subsection{Halo Merger Tree Representation}

\begin{figure}
\centering
\includegraphics[width=0.85\linewidth]{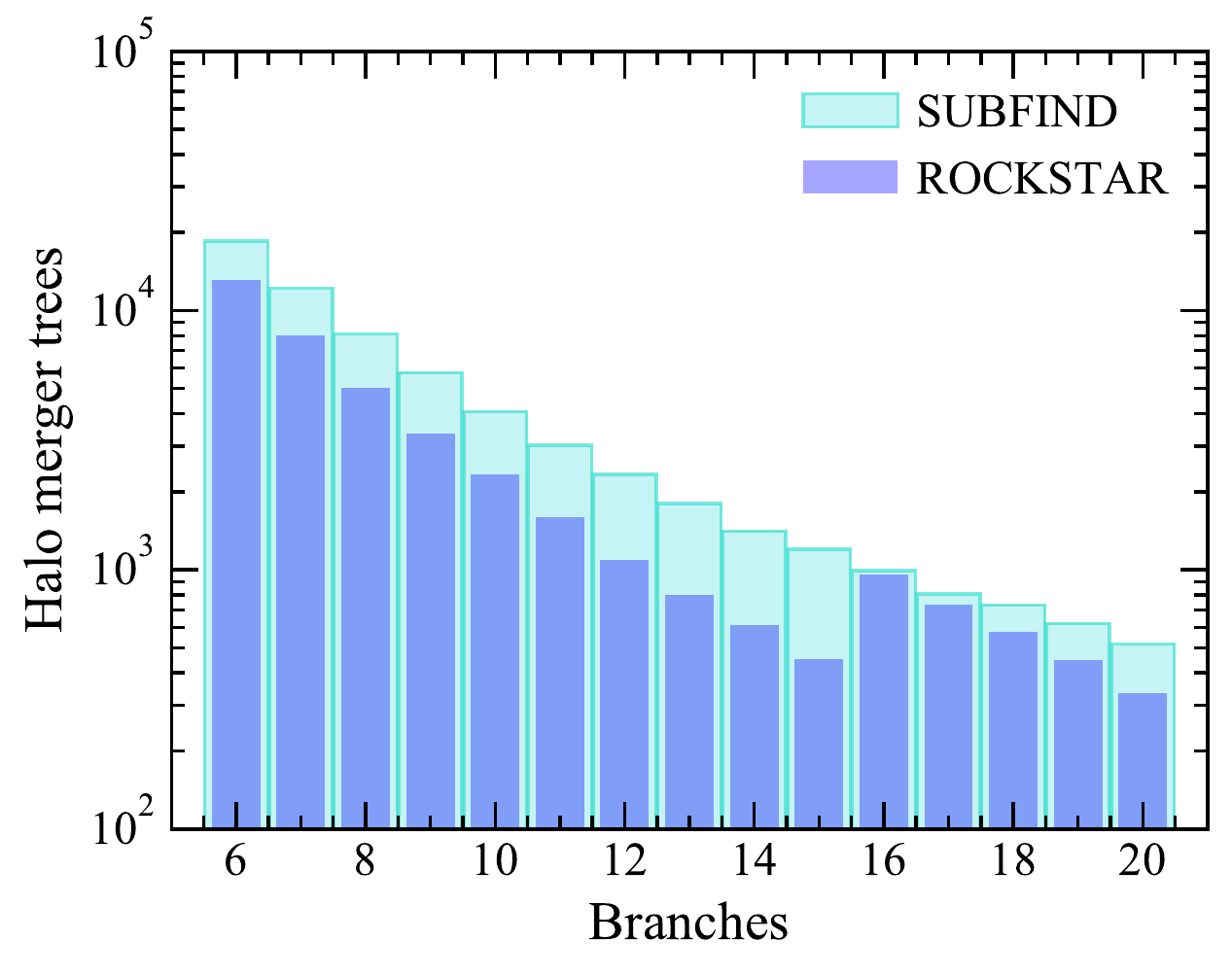}
\includegraphics[width=0.85\linewidth]{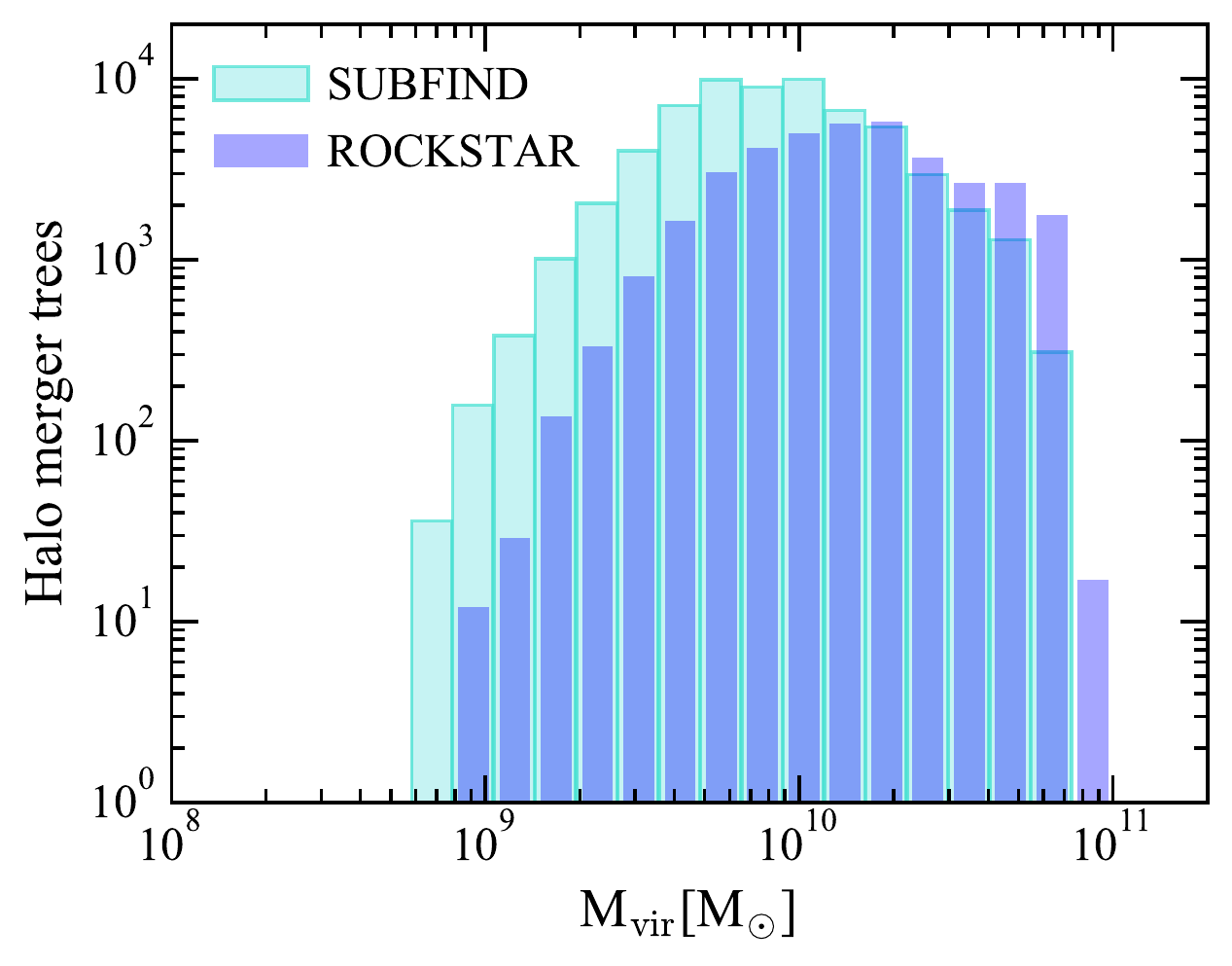}
\includegraphics[width=0.85\linewidth]{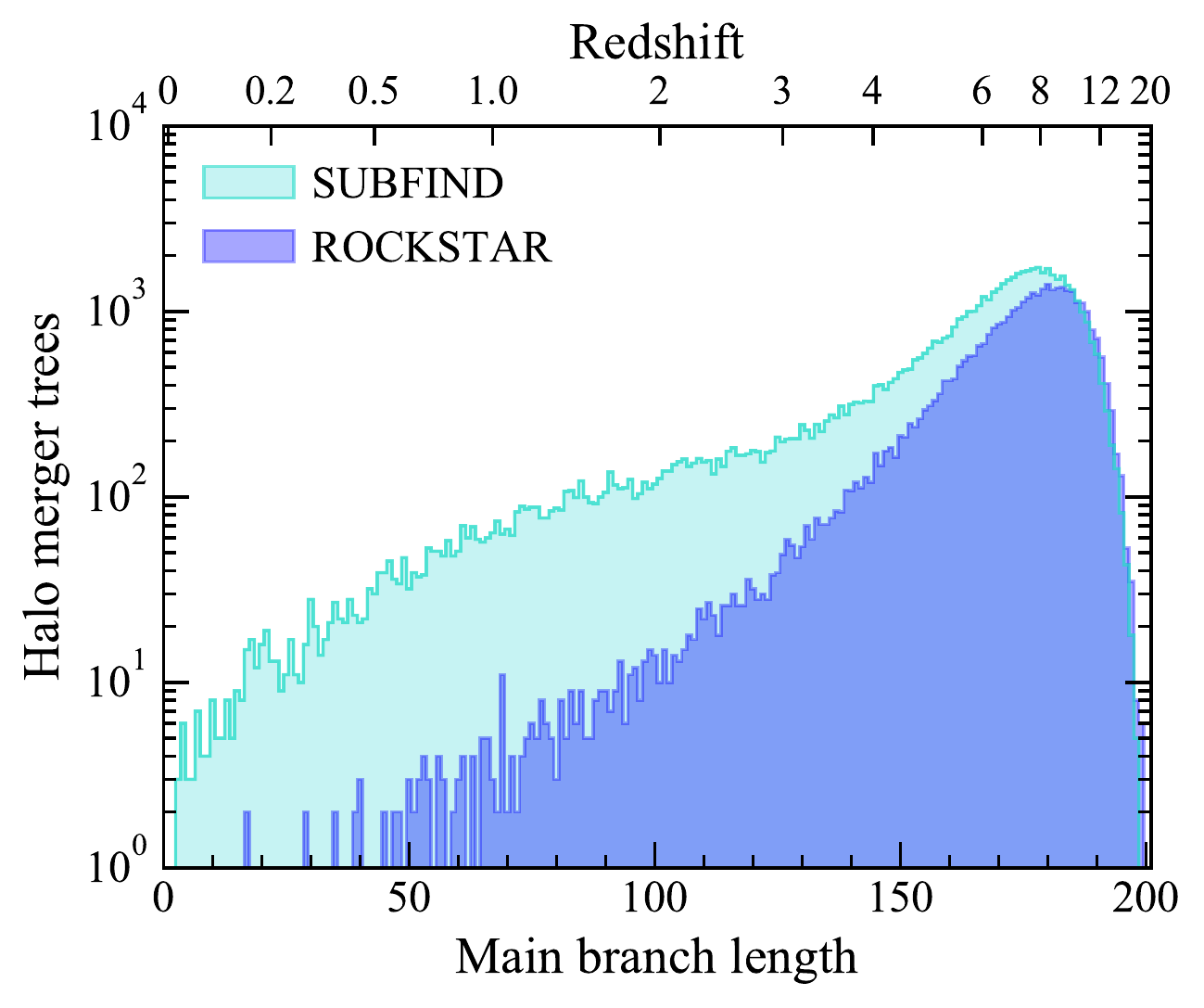}
\caption{Histograms of the merger trees in the training dataset. These are trees for main haloes at $z=0$,  identified using SUBFIND--D-TREES (light blue) and ROCKSTAR--{\sc ConsistentTrees} (blue), with  number of branches in the $6\leq\nbr\leq20$ range (top panel). 
This sample corresponds to main haloes with virial masses in the  $10^{9}\Msun \lesssim \Mvir \lesssim 10^{11}\Msun$ range at $z=0$ (middle panel).
The bottom panel shows the histogram of the length of the main branch measured in number of snapshots. (The redshift axis is also shown in the top axis for reference.)}
  \label{fig:traindataset}
\end{figure}

We selected halo merger trees for main haloes at $z=0$ from the aforementioned EAGLE  simulation, identified with SUBFIND--D-TREES and ROCKSTAR--{\sc ConsistentTrees}. The selection criterion is based on the number of trees available per number of branches. We select merger trees with at least 6 branches, since we are interested in reproducing relatively complex tree structures. 
The maximum number of branches considered during training, as we shall see in Section~\ref{sec:results}, will depend on the halo finder--tree builder algorithm. 
In Fig.~\ref{fig:traindataset}, top panel, we show a histogram of the initially selected merger trees binned by the number of branches. A large number of examples is required for our neural network model to properly learn to represent these tree structures. 
This initial selection results in  merger trees of main haloes with virial masses in the $10^{9}\Msun \lesssim \Mvir \lesssim 10^{11}\Msun$ range at $z=0$ (see middle panel of Fig.~\ref{fig:traindataset}).  
More massive haloes have  more intricate merger histories and  less examples of them are available in simulations. Note that for less massive haloes we also have fewer merger trees, this is due to the fact that they can have a smaller number of branches and  we are selecting merger trees with no less than 6 branches. To show how far in time, the merger histories of the haloes in the middle panel  of Fig.~\ref{fig:traindataset} can be traced back, we plot the length of the main branch of their corresponding merger trees in the bottom panel, measured in number of snapshots, from $z=0$ till the endpoint of the main branch at high redshift. We immediately notice that there are few merger trees with short main branches. These, in general, correspond to the low-mass haloes in our sample, as expected from the bottom-up theory of structure formation. Most of the trees in the trraining dataset tend to have long branches, with their distributions peaking slightly above $z\simeq8$, for both SUBFIND and ROCKSTAR trees. We also note that ROCKSTAR merger trees tend to feature longer main branches, with $\sim90\%$ of trees having main branches that span more than 150 snapshots back in time, i.e. up to $z\gtrsim 4.5$; while for SUBFIND merger trees this fraction reduces to $~73\%$. 

For a consistent description of the merger trees and motivated by  SAMs, we consider at most three quantities to be learned by our neural network, which are the most basic input variables required by SAMs \citep{Cole2000,Benson2012,Croton2016,Cora2018}. Namely, these are the  mass of the progenitors, distance of each progenitor to the corresponding halo in the main branch, and progenitor type, a discrete variable indicating whether the progenitor is a main or a satellite halo. 
Each variable was stored in matrix format, one matrix per variable, where columns represent branches and rows snapshots. 
Matrix elements were filled following the structure of a merger tree, where the first column is the main branch, i.e., the largest branch and progenitors in every branch and snapshot are denoted by non-zero matrix elements. 

We obtained separate databases for both halo finder--tree builder algorithms, SUBFIND--D-TREES and ROCKSTAR--{\sc ConsistentTrees}, hereafter referred as SUBFIND and ROCKSTAR, respectively. As mentioned above, we restricted  these databases to merger trees with number of branches  $\nbr\geq6$, see Fig.~\ref{fig:traindataset}. These are among the most abundant merger trees  for haloes in the above mentioned mass range. 

\begin{figure*}
\centering
\vspace{10pt}
\includegraphics[width=0.75\linewidth]{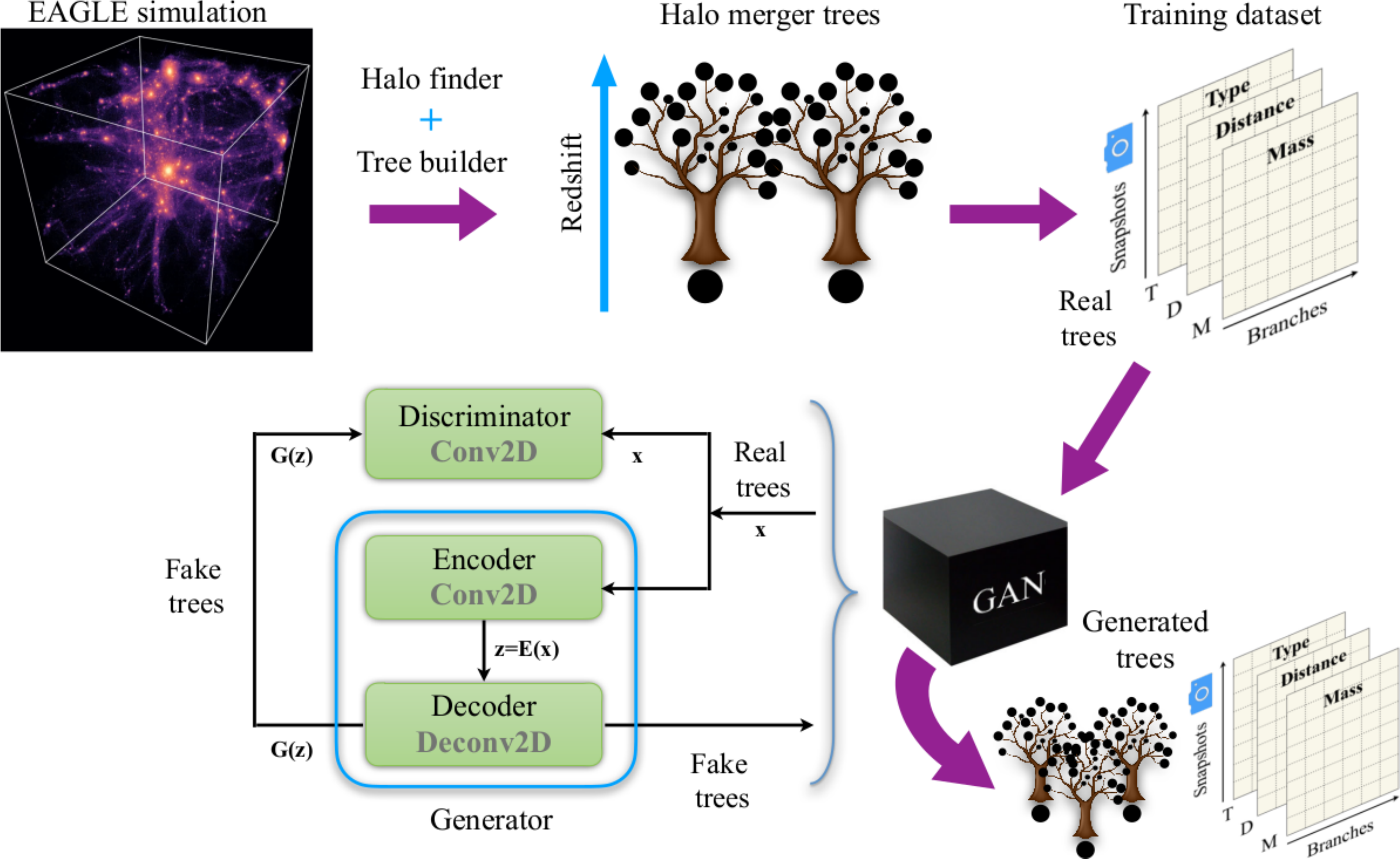}
\caption{%
Halo merger tree generation process \citep[adapted from][]{Robles:2019nfk}. Halo merger trees from the DM only EAGLE simulation of 100 cMpc constructed with either SUBFIND--D-TREES or ROCKSTAR--{\sc ConsistentTrees} are stored in matrix format (one matrix per input variable), forming two separate datasets. Batches of these `real' merger trees  are the inputs of our GAN model comprised of a discriminator and a generator that consists of an encoder and a decoder, all these neural networks are made of convolutional layers (Conv2D, Deconv2D). The outputs of the GAN are batches of `fake' (generated) trees in matrix format, as many matrices as input channels are obtained. 
} 
\label{fig:mt_generation}
\end{figure*}

\subsection{Neural Network Model}
\label{sec:GANmodel}

Generative Adversarial Networks are a framework designed to learn generative models of complex data distributions. GANs consist of two neural networks a generator and a discriminator in competition with each other. 
Their ultimate goal is that the generator learns a distribution that matches the real data, while fooling the discriminator that is trained to distinguish real from generated data samples.  

The GAN architecture adopted here is based on the layout previously designed and tested by \citet{Robles:2019nfk} to generate halo merger trees with a fixed number of branches, namely $\nbr=6$, and a temporal resolution of $29$ snapshots. In this model, the generator comprises an encoder-decoder architecture \citep{Bengio2013} that learns to reproduce the matrix representation of merger trees.
Both, the discriminator and the encoder are implemented with convolutional neural networks (CNNs) \citep{Krizhevsky2012}, while the decoder is made of deconvolutional layers. Each CNN layer features either row- or  column-wise filters, which intend to reproduce operations within a branch (column-wise filters) and among merging branches (row-wise filters). 
Finally, reconstruction losses for each considered input (typically the mass of the progenitors and the final descendant) are added to the classic GAN loss function. These reconstruction losses drive the learning process and improve considerably the quality of the generated trees. All the loss functions are computed with cross-entropy measures. 
Both the discriminator and the generator are trained with Adam, a stochastic gradient-based optimisation algorithm \citep{Kingma:2014}, with batches of `real' merger trees in matrix format. 

In this paper, we generalise the GAN architecture introduced by \citet{Robles:2019nfk} to produce merger trees with higher time resolution which results in some of the CNN layers featuring filters with larger kernel sizes; the precise size depends on the halo finder--tree builder algorithm employed to construct the merger trees in the training database and the maximum number of branches considered. We also remove the restriction on the number of branches, namely we train our model with merger trees with higher number of branches, $\nbr\geq6$, fixed and arbitrary in a given range. These modifications also alter other parameters of the GAN model such as the size of the input of the decoder and the batch size for training. The exact details of the GAN architecture are given in Appendix~\ref{sec:GANarchitecture}. It is worth remarking that the maximum number of branches that our GAN model can learn to generate is limited by the amount of merger trees in the training dataset and memory resources. We noticed that around $1000$ merger trees per given $\nbr$ are required for the GAN total loss to converge. This number of merger trees  is easy to obtain from any DM-Only simulation nowadays, depending on the  mass of the final descendant and the volume of the simulation. 
As can be seen from Fig.~\ref{fig:traindataset}, this imposes a more stringent restriction on the maximum number of branches of a tree that can be reconstructed using merger trees of the E100 simulation identified with ROCKSTAR.

\begin{figure*}
\centering
\includegraphics[width=\columnwidth]{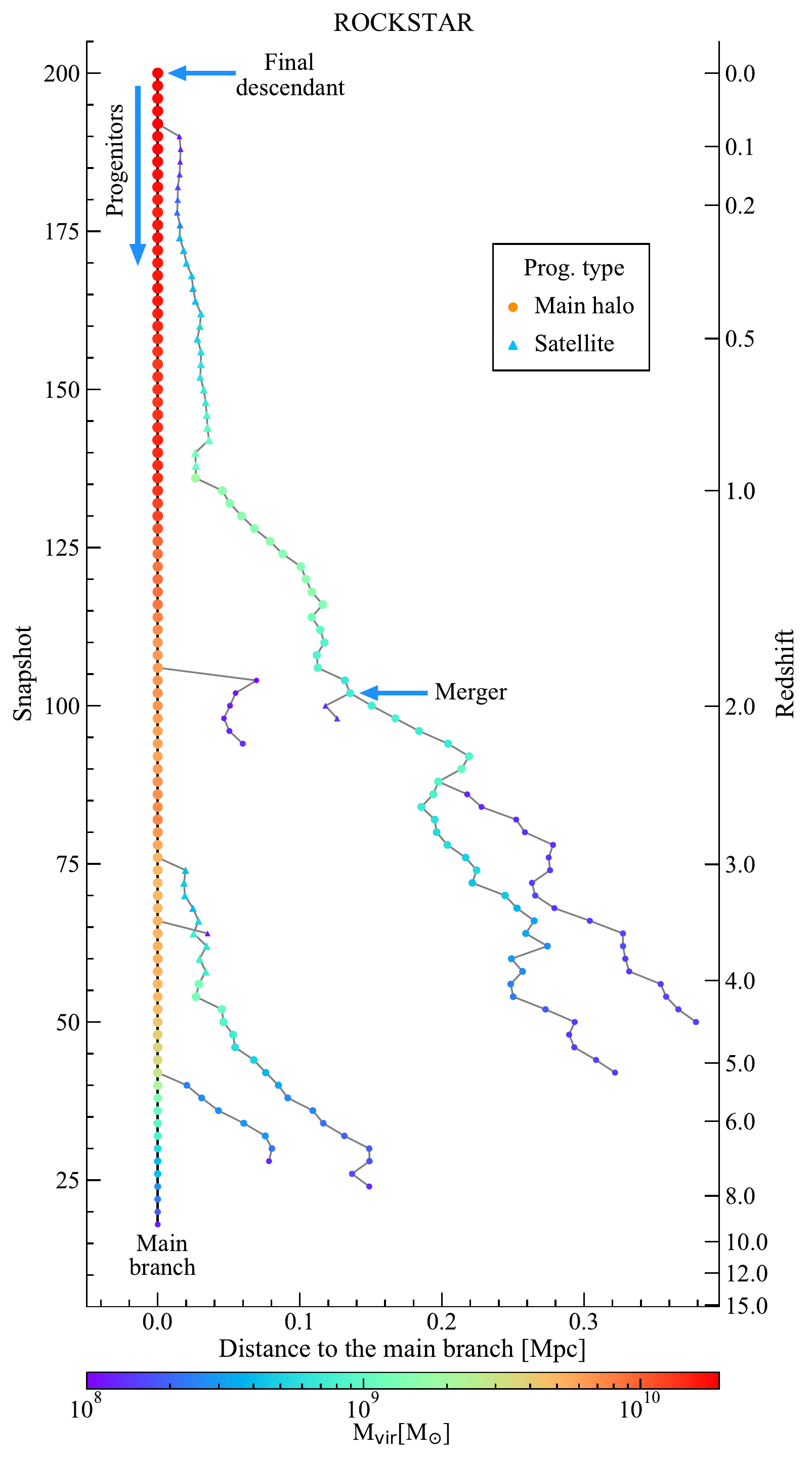}
\includegraphics[width=\columnwidth]{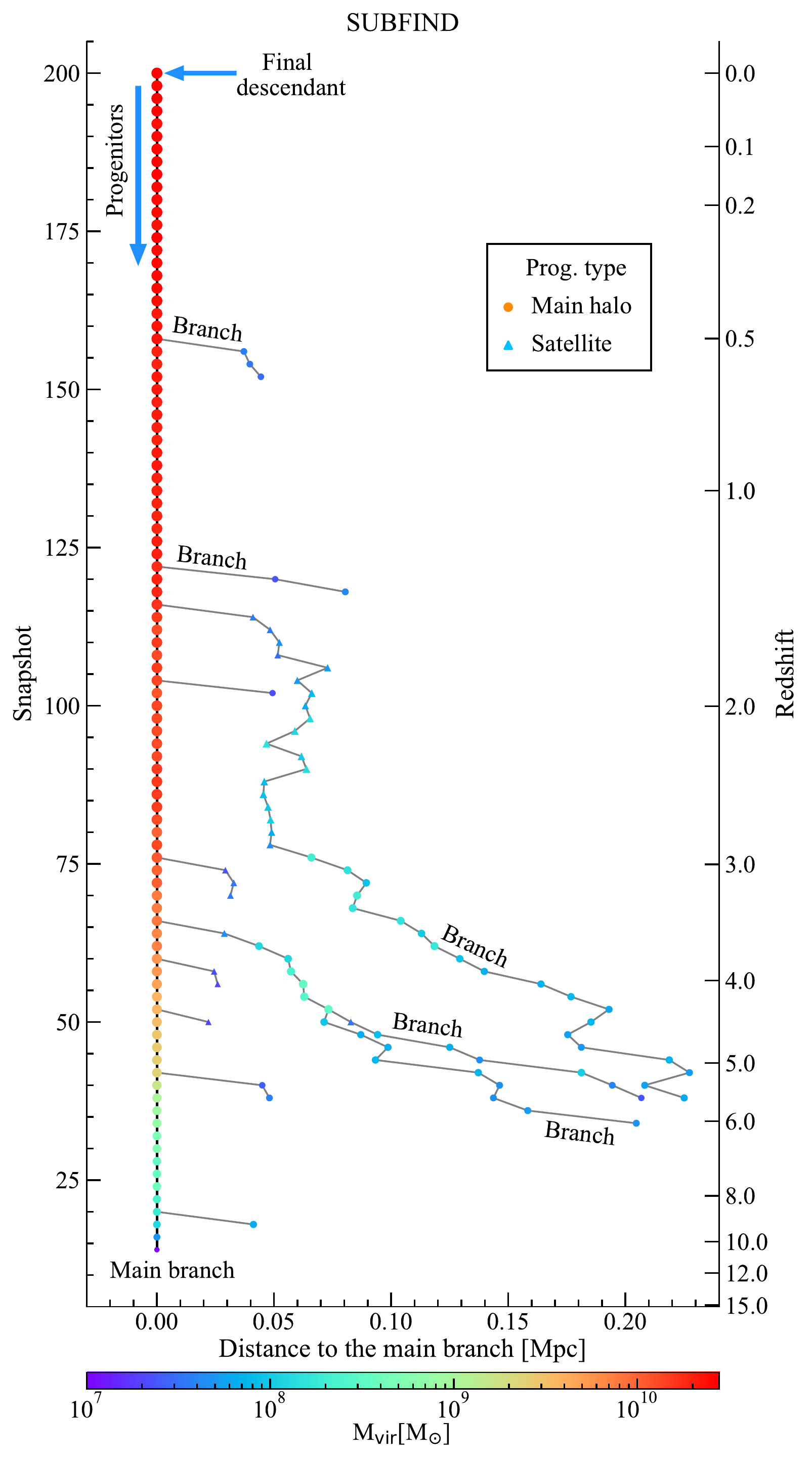} 
\caption{
  Examples of ROCKSTAR-like (left-hand panel, 8 branches) and SUBFIND-like (right-hand panel, 12 branches) machine learning generated merger trees   in the plane snapshot (redshift) vs.\ distance to the main branch. 
  Progenitors that are main haloes are denoted by  circles and satellites by triangles,  the colour map represents the virial mass of the halo progenitors. 
 }
  \label{fig:mergertree}
\end{figure*}

This refurbished GAN model is trained with a dataset of `real' merger trees from the E100 simulation, identified with either SUBFIND or ROCKSTAR, where each considered variable (up to three: mass of the progenitors and the final descendant, distance to the main branch, and progenitor type) corresponds to an input channel. 
As in any unsupervised learning technique, only examples of the inputs are given to the GAN, which learns to reproduce them. 
Thus, when adding more properties of the progenitors of a halo in a merger tree  structure, the GAN model must also learn to generate this information. This is reinforced by the addition of  a reconstruction loss per additional input.  
As a result, the GAN yields as many outputs as input channels per generated merger tree. 
Note that in principle, it is possible to add as many extra inputs as desired, subject to memory resources. 
The complete generation process is summarised in Fig.~\ref{fig:mt_generation}. The training process is performed in batches of randomly shuffled merger trees, several epochs after the total GAN loss converges, we obtain as many batches of  generated (`fake') trees as convenient for the analyses in Section~\ref{sec:results}.
Examples of these ML generated trees are shown in Figs.~\ref{fig:mergertree} and \ref{fig:mergertreestruct}.\footnote{For visualisation purposes, we show merger trees generated with $\nsnaps=101$, time resolution that as we shall see yields the best results.} 

\subsection{Evaluation of Generated Merger Trees}
\label{sec:evaluation}

Our neural network learns to generate well-constructed merger trees, i.e., progenitors with no drastic variation in their mass and no sudden jumps in physical location. 
To quantitatively evaluate the quality of our GAN-generated trees, as in \citet{Robles:2019nfk}, 
we construct probability distributions of the training dataset and equally sized samples of the reconstructed trees for a given total number of branches, a given time resolution (i.e. number of snapshots: $\nsnaps=201$ or $\nsnaps=101$) and for each considered variable. We later compare these distributions using the Kolmogorov-Smirnov (KS) test.

\begin{figure}
    \centering
    \includegraphics[width=0.9\linewidth]{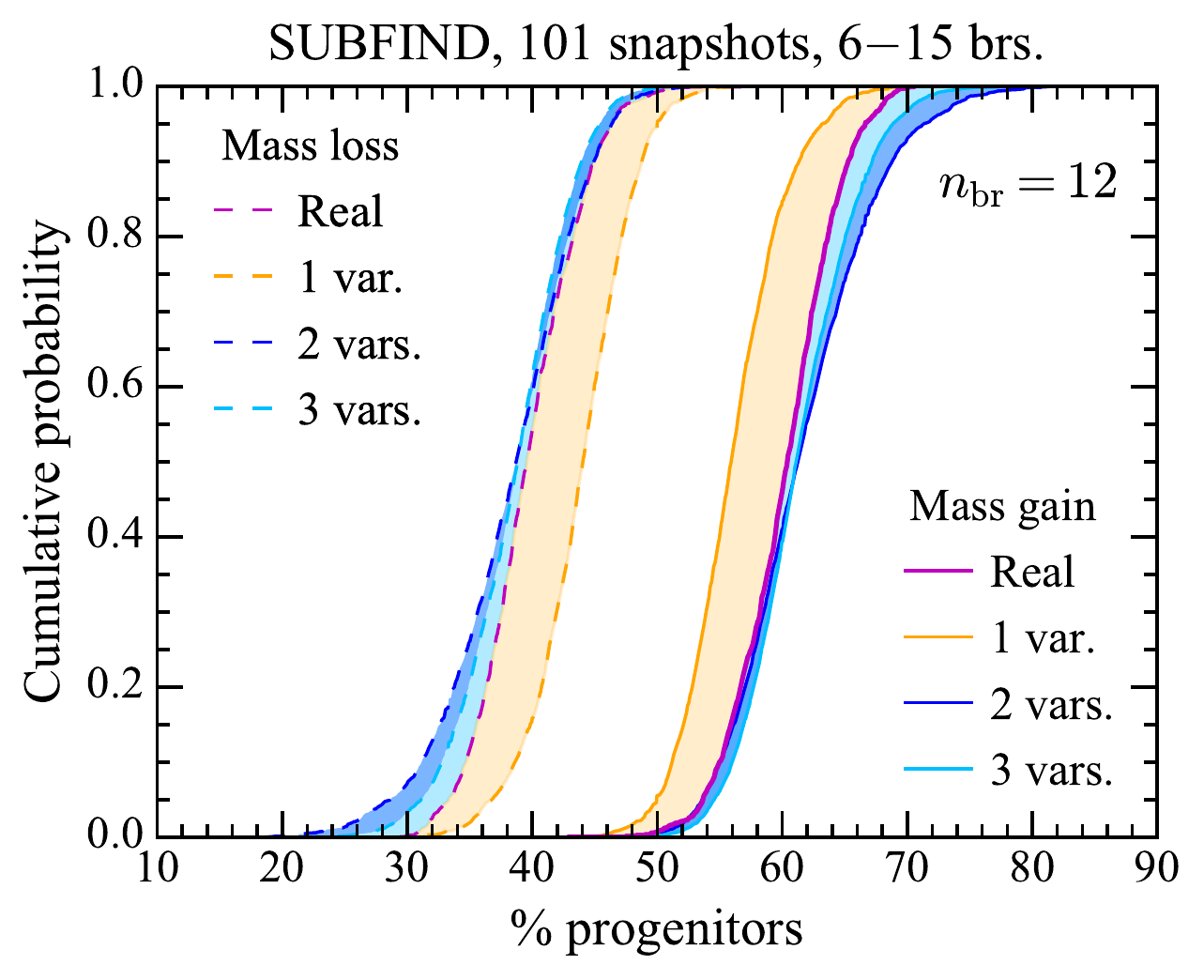}
    \caption{Cumulative distribution function (CDF) of the mass gain (solid) and loss (dashed) for all the progenitors in the SUBFIND merger tree dataset with 12 branches, denoted `real' (magenta). For comparison, we also show the CDFs for samples of trees generated with 1 (orange), 2 (blue), and 3 (light blue)~variables, time resolution of 101 snapshots, and the GAN model trained with a dataset composed of trees with $6\leq\nbr\leq15$. The shaded regions depict the difference between the `real' and ML  CDFs. 
   }
    \label{fig:metric_full_mass}
\end{figure}

It can be argued that the most important information a merger tree provides is the mass assembly history of a halo, i.e. the mass of their progenitors.  The mass growth is the main input in galaxy formation models (e.g. \citealt{Springel2001}). Masses along a branch of a tree  are expected to monotonically increase as time elapses, especially for main haloes. Nevertheless, as we shall see this behaviour, assumed in SAMs \citep{Lacey1991, White1991, Cole1994, Cole2000, Gonzalez-Perez2014, Lacey2016, Lagos2018}, in fact depends on other properties of the progenitors,  such as being a main or a satellite halo, and  physical quantities, e.g. the distance between merging progenitors. Therefore, we do not enforce during the training process a progenitor mass increasing with time in a branch. To evaluate the fair reproduction of the mass growth in a sample of ML-generated  trees, we compare statistical distributions of the mass gain and loss of progenitors for merger trees with fixed number of branches. We show an example of these distributions in  Fig.~\ref{fig:metric_full_mass}, where `real' denotes the cumulative probability  obtained using merger trees from the training dataset. For a description of the combination of variables considered in the ML distributions see Table~\ref{tab:trainings}. 
From Fig.~\ref{fig:metric_full_mass}, we immediately note that there is an improvement in the learning process when providing the GAN model with more information of the merging events at each time-step, i.e., by 
including relevant variables or inputs (see shaded regions). 

\begin{figure}
\centering
\includegraphics[width=\linewidth]{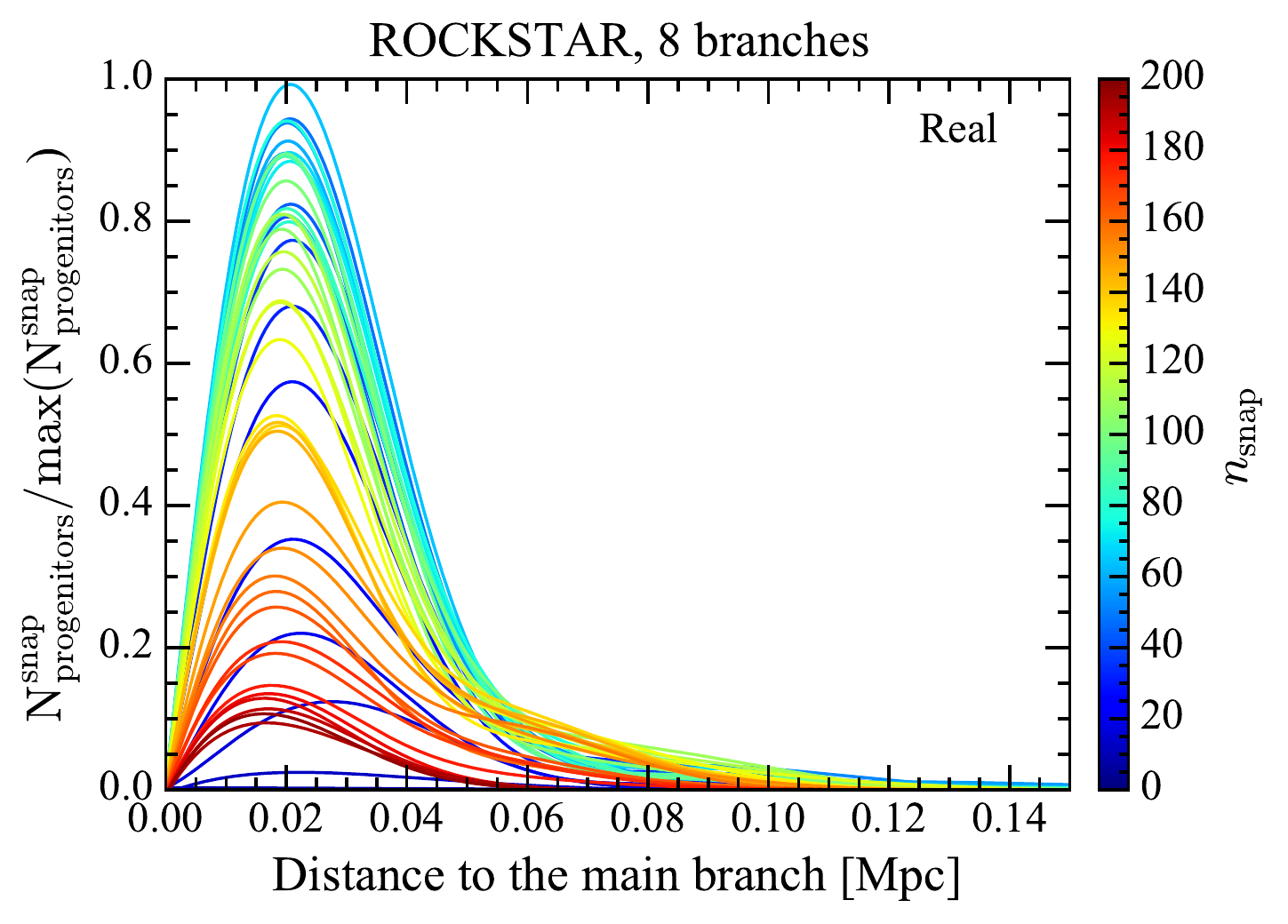}
\caption{%
  Distribution of the distance between merging progenitors, when one of them is located in the main branch,  for merger trees with 8 branches identified by ROCKSTAR (denoted `real'). 
  The distance is computed at the snapshot before the merging event takes place  (colourmap). 
  The number of progenitors at a given step in time $\mathrm{N_{progenitors}^{snap}}$ is normalised by the maximum peak of all distributions. 
}
\label{fig:distancedistrib}
\end{figure}

Another input of our GAN model is the physical distance between merging haloes, specifically we consider the distance between the centre of mass of progenitors in branches other than the main branch and that in the main branch.  It should be noted that most semi-analytic models of galaxy formation do not require this variable to evolve galaxies (e.g. \citealt{Somerville2008}). Being able to predict this quantity would allow further physics to be included in such models; as in the case of GALFORM~\citep{Cole2000}, a SAM that takes into account these distances (together with other physical parameters) to redefine the  progenitor type (main or satellite halo; a variable that we have also considered) of a infalling halo \citep{Helly2003,Jiang2014}. 
In principle, the distance between progenitors should decrease with time as the merging event approaches, but there is no rule of thumb for the precise step in time this should occur. Hence, we construct normalised probability distributions of this distance at the snapshot before the fusion takes place, for real and ML  merger trees with the same number of branches (see, e.g., Fig.~\ref{fig:distancedistrib}. Note that the precise  snapshot at which a merger occurs varies  from $0$ (dark blue) up to the last snapshot, $200$ (dark red), as shown in   Fig.~\ref{fig:distancedistrib}, where we have normalised the distributions by the maximum peak of all curves, to facilitate comparison. In this figure, we observe that as expected, as time elapses merging haloes are found at a closer distance from each other one snapshot before the fusion. Hence, the peak of the distributions from $\nsnaps=0$ to $200$  progressively shifts to the left. 
We can also observe that most mergers occur at $\nsnaps=60$ ($z\simeq3.8$) for ROCKSTAR merger trees with 8 branches. 
With the above mentioned distributions, properly normalised, we construct cumulative distribution functions (CDFs), one distribution per snapshot    (curves in  Fig.~\ref{fig:distancedistrib}), for the real and ML merger tree samples. These CDFs are suitable to compare using the KS test. 

\begin{figure}
    \centering
    \includegraphics[width=0.9\linewidth]{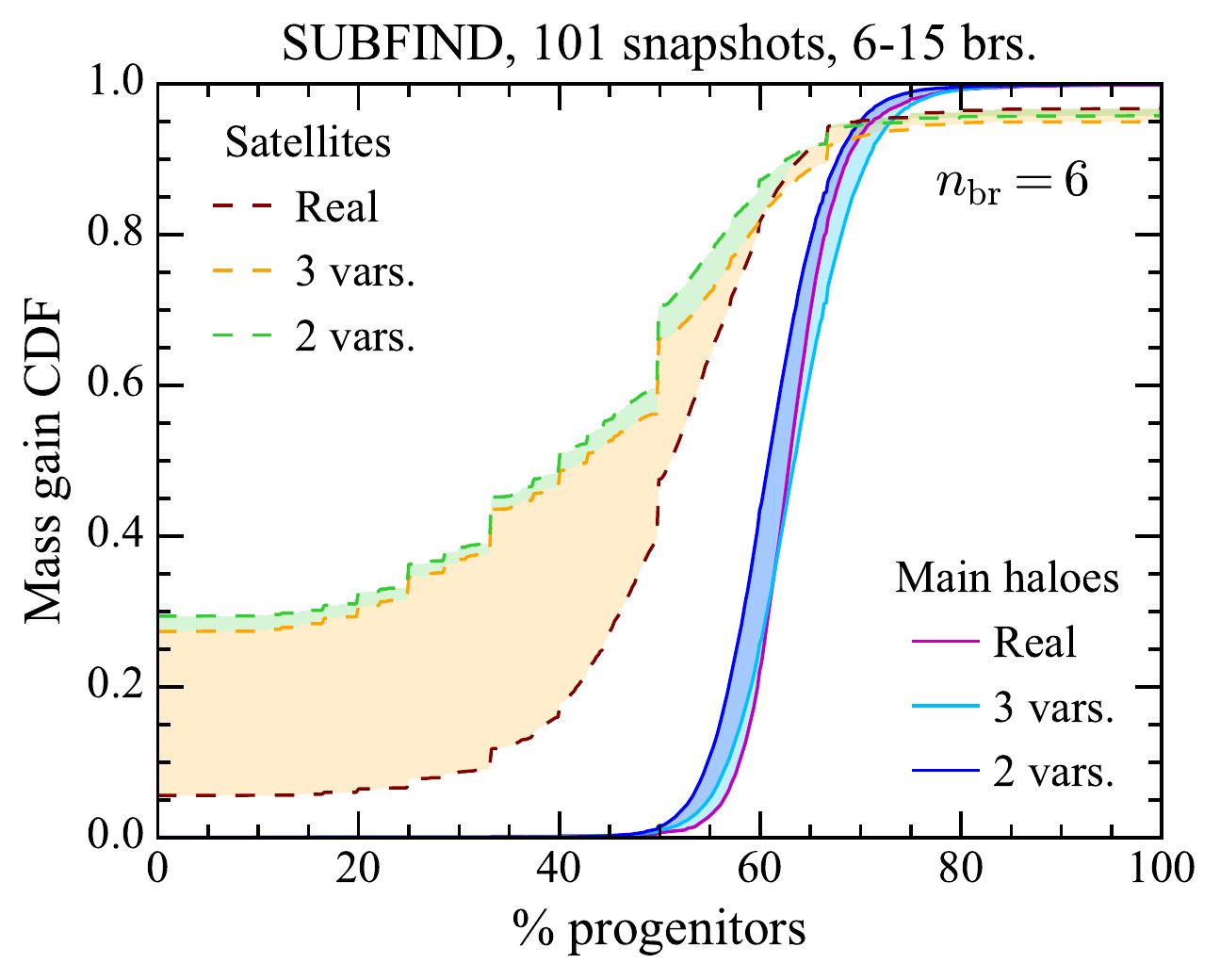}
    \caption{CDF of the mass gain for all the progenitors that are main haloes (solid) and satellites (dashed)  in the SUBFIND merger tree dataset with 6 branches, denoted `real' (magenta). For comparison, we also show the CDFs for samples of trees generated with  2 (mass and progenitor type) and 3~variables. ML trees generated with a time resolution of 101 snapshots and  a training dataset composed of trees with number of branches in the range $6\leq\nbr\leq15$. The shaded regions highlight the difference between the `real' and ML  CDFs. 
   }
    \label{fig:metric_partial_mass}
\end{figure}

\begin{figure}
\centering
 \includegraphics[width=0.9\linewidth]{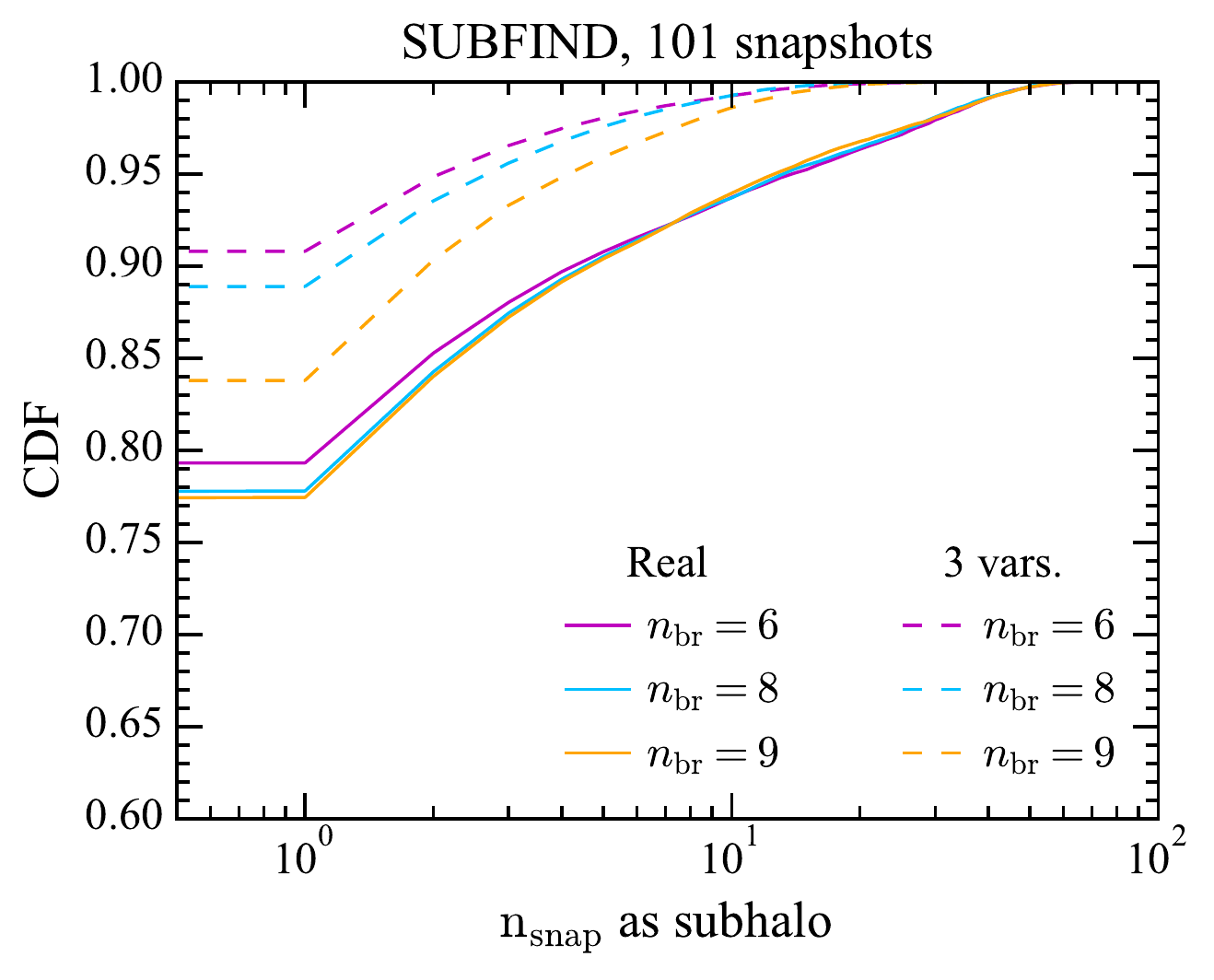}
    \caption{CDF of the number of snapshots that a progenitor is a satellite halo for the real and ML  trees with $\nbr=6,8,9$, generated with 3 variables (individual trainings with a time resolution of $\nsnaps=101$ were performed for every $\nbr$ in this figure).} 
    \label{fig:N_sns_as_sh}
\end{figure}

The third variable that we have considered is the progenitor type, i.e., the condition of being a main or a satellite halo. This variable is fundamental in semi-analytic models of galaxy formation, such as GALFORM~\citep{Cole2000}, SAGE~\citep{Croton2016}, SAG~\citep{Cora2018} and GALACTICUS~\citep{Benson2012}, among others; as it  strongly influences the physical processes that take place in a galaxy, such as cooling (see e.g. \citealt{Cole2000}), affecting the cold gas component and the star formation rate (see e.g. \citealt{Gomez2021}). Halo progenitors in the main branch, with the sole exception of early snapshots, are expected to be main haloes. Conversely, progenitors in other branches can become satellites as a consequence of gravitational infall as they approach the other merging halo. The precise step in time when this occurs depends on the specific tree and also on the  masses and distance of the merging progenitors \citep{Diemand2006, Muldrew2011, Han2012, Onions2012, Elahi2011, Onions2013}. Satellite haloes approach the main haloes they are going to merge with and their mass is allowed to decrease in time as the main halo grows in mass. An example of this behaviour can be found in Fig.~\ref{fig:mergertree}. Therefore, we can evaluate the importance of including the progenitor type when reproducing its  corresponding mass, by analysing  the above mentioned mass gain and loss of the progenitors, but splitting each real and ML merger tree sample in two sub-samples according to  the progenitor type, as illustrated in Fig.~\ref{fig:metric_partial_mass}. Then, we proceed to compare real with ML distributions for main and satellite haloes. For that particular example, we can see that the mass gain (and loss) of the main haloes is much better reproduced by the GAN model than that of the satellites. This is probably due to the fact that most of the progenitors in a merger tree are main haloes. 
In this sense, we can perform an additional test and construct distributions of the number of snapshots a progenitor spends as a satellite, see, e.g.,  Fig.~\ref{fig:N_sns_as_sh}. This is an explicit test of the fair reconstruction of the progenitor type input. When comparing this figure with Fig.~5 of \citet{Robles:2019nfk}, we note that when adding more time resolution, it becomes more difficult for the GAN model to predict in which time-step a progenitor becomes a satellite. 

\section{Results}
\label{sec:results}

\begin{table*}
    \centering
    \caption{Series of trainings performed using either SUBFIND or ROCKSTAR halo merger trees, $\nsnaps$ denotes the number of snapshots (time resolution) and $\nvar$ the number of input variables, where 1 var. refers to the mass of the progenitors and the final descendant, hereafter mass, 2 vars.\ corresponds to trainings with either mass and distance to the main branch or mass and type of progenitor, 3 vars.\ refers to all the aforementioned inputs.}    
    \begin{tabular}{ccccc}
    \hline
   Training & $\nvar$ & $\nsnaps$ & \multicolumn{2}{c}{Number of branches $\nbr$} \\
   \cline{4-5}
    &  & &  SUBFIND & ROCKSTAR \\
    \hline
    \multirow{2}{*}{single $\nbr$} & 1 var. & 101, 201 & 6--15 & 6--13 \\    
     & 2--3 vars. & 101 & 6--15 & 6--13 \\\hline 
    \multirow{3}{*}{multiple $\nbr$} & 1 var. & 101, 201 & 6--10, 6--12, 6--15 &  6--10, 6--12, 6--13 \\
     & 2 vars. & 101 & 6--10, 6--12, 6--15 &  6--10, 6--12, 6--13 \\   
     & 3 vars. & 101 & 6--10, 6--15, 6--16, 6--19 &  6--10, 6--12, 6--13 \\       
    \hline
    \end{tabular}
    \label{tab:trainings}
\end{table*}

In this section, we apply the statistical measures outlined in the previous section and compare ML  with real distributions using the Kolmogorov-Smirnov test. First of all, in Table~\ref{tab:trainings} we summarise the series of trainings we have performed to study the relevance of each of the variables considered  here to the well construction of the mass growth history of haloes, and other aspects related to the performance of our GAN model. These aspects include the quality of the  merger trees generated with individual trainings, i.e., with datasets comprised of merger trees with a fixed number of branches denoted single $\nbr$ and trainings with $\nbr$ varying in a range (multiple $\nbr$). Recall that the number of branches is a fundamental parameter of the architecture of our GAN model, see Table~\ref{tab:GANarch}; so that for our series of experiments,  we select sub-samples of the initial training dataset in Fig~\ref{fig:traindataset}   based on $\nbr$ and not on the final descendant mass, to maximise the amount of examples of different tree structures per given $\nbr$ and hence to ensure the convergence of the training process. 
In this sense,
multiple $\nbr$ trainings are a more realistic scenario than single $\nbr$ since a set of merger trees  with different numbers of branches are generated at once for final descendants with masses than span a wider range, as in cosmological simulations.

\subsection{Mass assembly history}
\label{sec:mass}

\begin{figure*}
    \centering
    \includegraphics[width=0.77\linewidth]{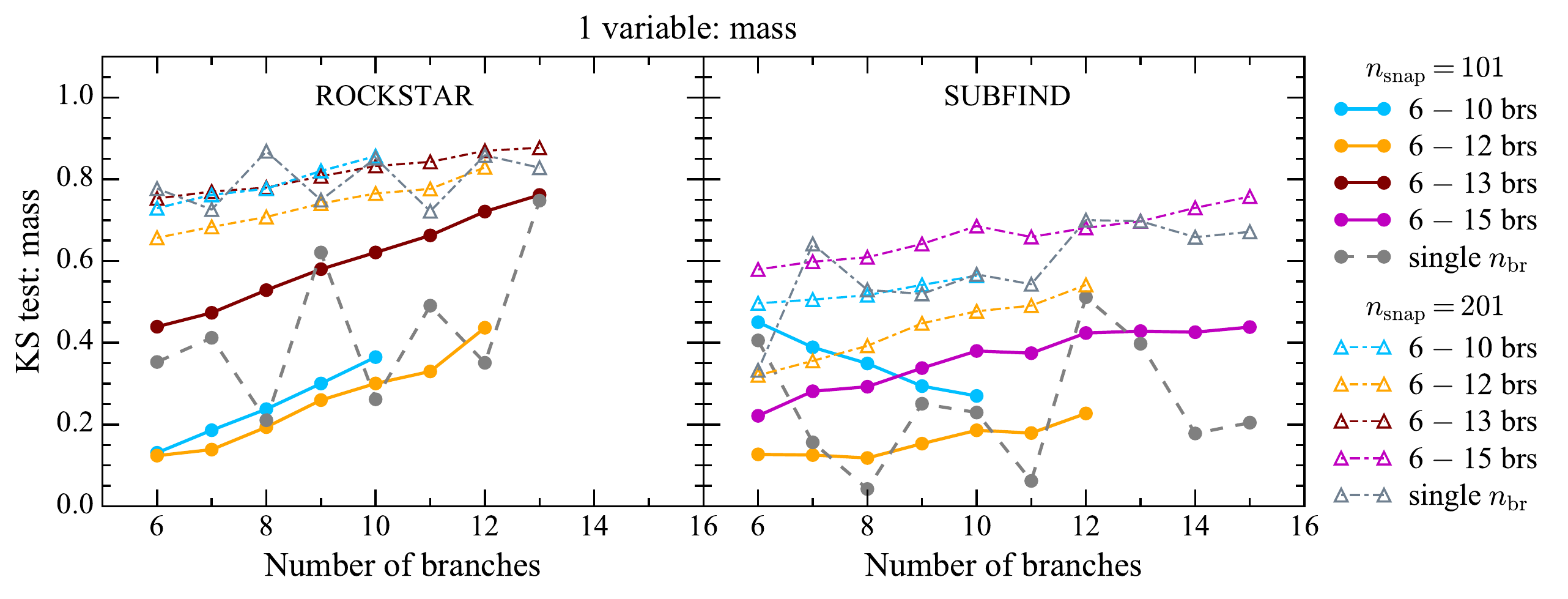}
    \includegraphics[width=0.77\linewidth]{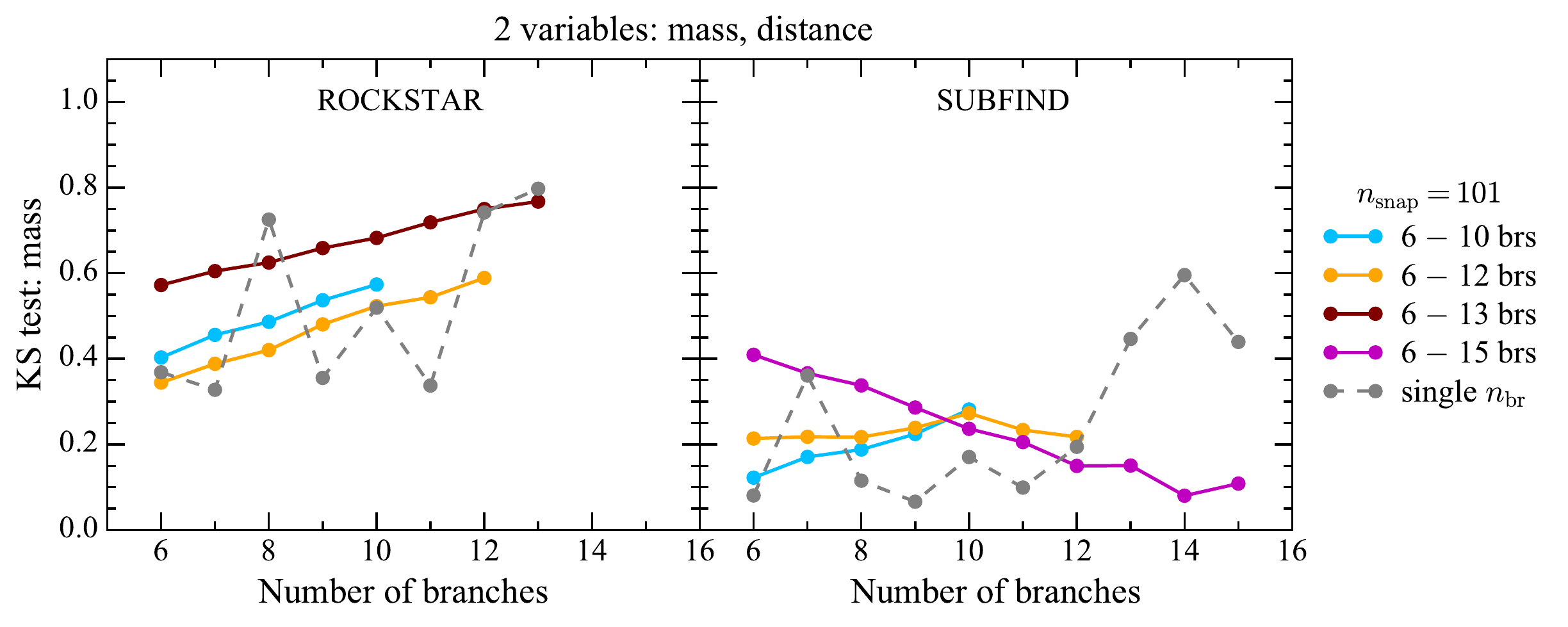}    
    \includegraphics[width=0.77\linewidth]{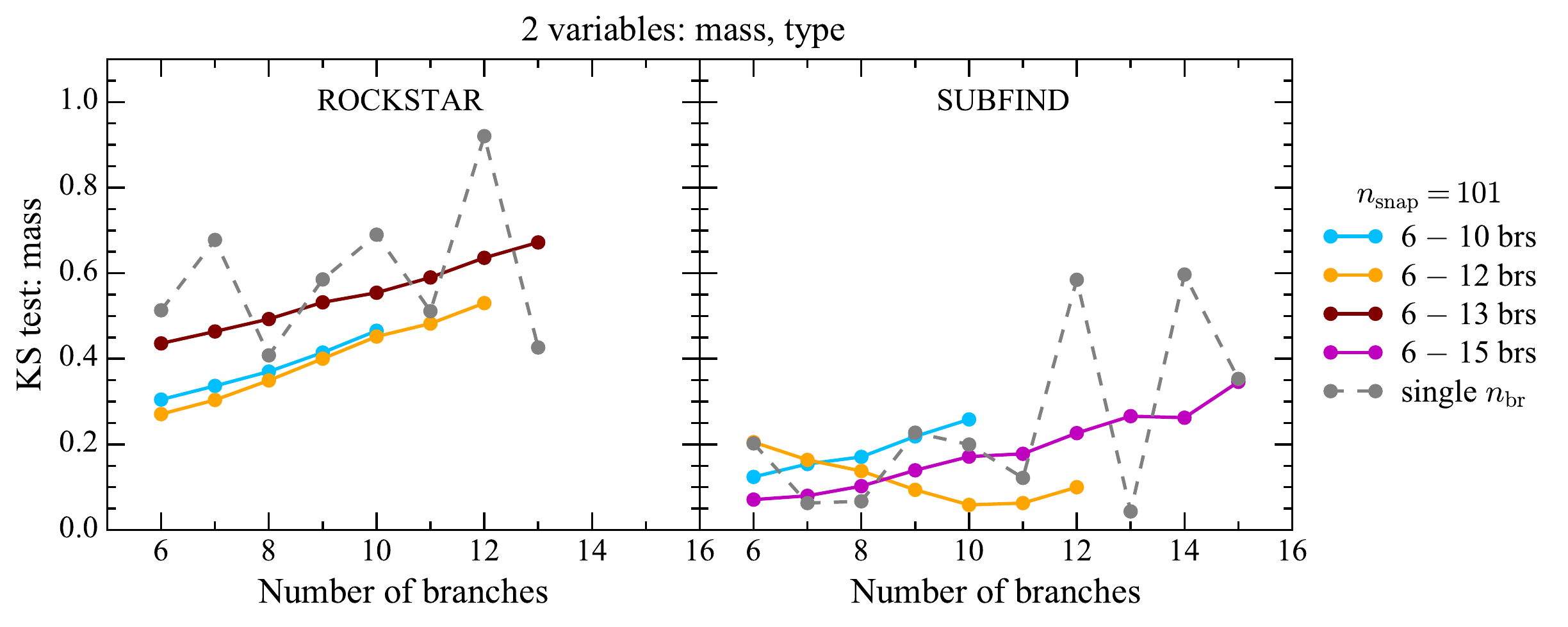}    
    \includegraphics[width=0.77\linewidth]{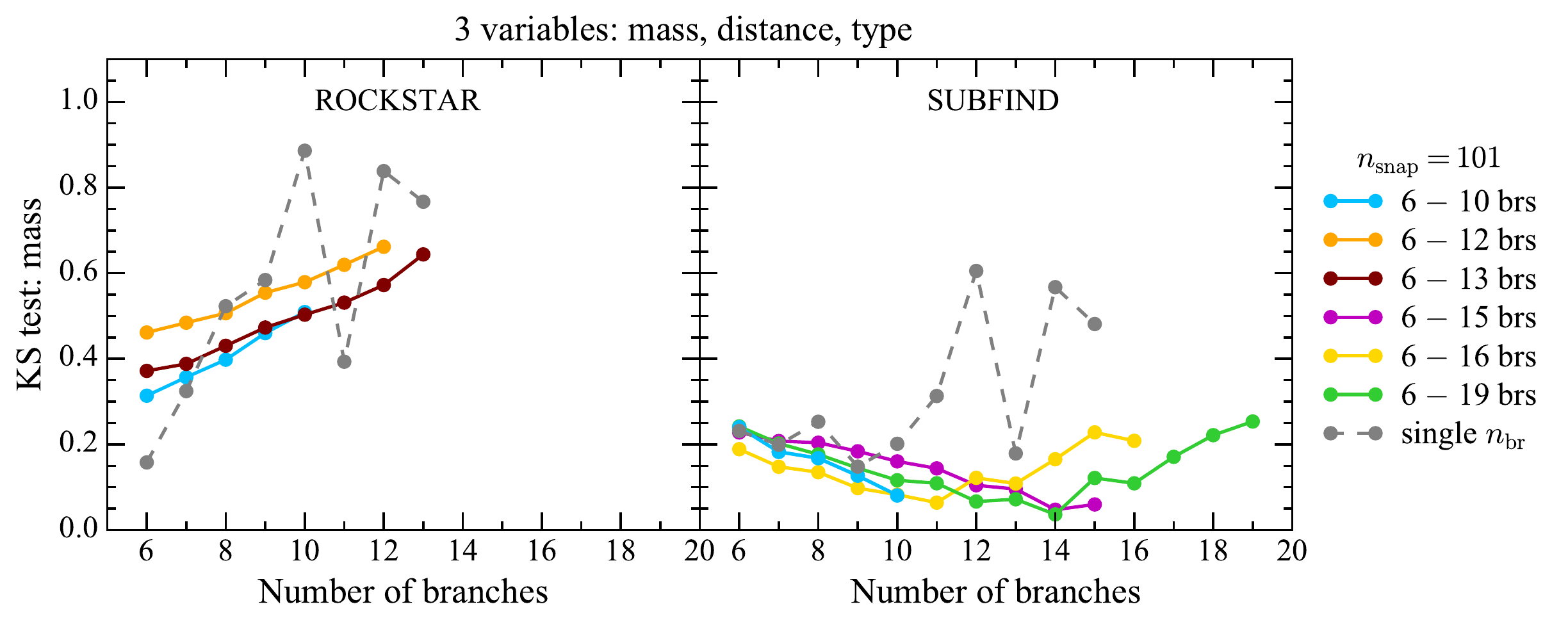}        
    \caption{Kolmogorov-Smirnov (KS) test for the mass of the progenitors and the final descendant for merger trees generated with 1 (top panels), 2 (second row: mass and distance to the main branch, third row: mass and progenitor type) and 3 (bottom panels) input variables, GAN model trained with either ROCKSTAR (left-hand panels) or SUBFIND (right-hand panels) merger trees. Grey lines correspond to series of individual trainings (single $\nbr$, one training per $\nbr$) and colourful lines to single trainings performed with  trees whose number of branches vary in a range (multiple $\nbr$, a single training per line). }
    \label{fig:KStestmass}
\end{figure*}

First, we focus on the fair reproduction of the mass assembly history of haloes and train our GAN model considering the progenitor mass as the sole input. 
Initially, we constructed training datasets with the maximum achievable resolution in time, i.e., $\nsnaps=201$. 
In Fig.~\ref{fig:KStestmass}, top panels we show the results of the KS test from the comparison of the mass gain and loss cumulative distributions (for an example of these CDFs, see Fig.~\ref{fig:metric_full_mass}) as dot-dashed lines. Recall that we construct CDFs for `real' and ML merger trees with a given $\nbr$, regardless if the training was performed with a single or a multiple $\nbr$ dataset. 
We carried out four series of trainings for each halo finder--tree builder algorithm (denoted SUBFIND and ROCKSTAR), three of them using merger trees with  number of branches in the  6--10 (light blue), 6--12 (orange), 6--13 (brown) and 6--15 (magenta) ranges. 
Note that the maximum number of branches we have considered varies with the algorithm, $\nbr=13$ for ROCKSTAR and $\nbr=15$ for SUBFIND\@. This is due to the size of the training dataset, namely  ROCKSTAR finds only $\sim800$ trees with 13 branches, while for SUBFIND there are $\sim 1800$ trees with this number of branches in the training dataset,  see Fig.~\ref{fig:traindataset}.  
As mentioned in the previous section, $\sim1000$  trees for a given $\nbr$ are required to obtain well-constructed merger trees. 
The grey lines denoted  `single $\nbr$' represent individual trainings, i.e., every point corresponds to a  training with fixed $\nbr$. E.g., for SUBFIND the grey lines represent 10 independent trainings, while each  magenta line was obtained with a single training. 
We immediately notice that the learned representation of SUBFIND merger trees  always  gives better results than that of ROCKSTAR\@. Note however that there are  more merger trees identified with SUBFIND in the training database than ROCKSTAR examples for our GAN to learn to represent (see Fig.~\ref{fig:traindataset}), and the morphology of SUBFIND and ROCKSTAR merger trees is not necessarily similar for haloes of the same mass.

As expected, reducing the temporal resolution  to $\nsnaps=101$, distributed between $z=20$ and $z=0$, allows us to improve the quality of the generated trees. 
Note that in general better values of the KS test are obtained with datasets that include trees with number of branches within a range rather than individual trainings with fixed $\nbr$, because of the larger size of the dataset. This is more evident for trainings with $\nsnaps=101$, compare dashed grey lines with solid lines. 
In both cases $\nsnaps=201$ and $\nsnaps=101$, the learned representation of the mass of the progenitors that better fits the real mass gain and loss distributions  is that obtained with the $6\leq\nbr\leq12$ dataset for both SUBFIND and ROCKSTAR ML trees\@. Of course, increasing the maximum number of branches in the training dataset enlarges its size and in principle improves the learning process. Although, the GAN model would have to  learn to reproduce a more complex tree structure with less available examples, since in general there are less trees in the E100 simulation with larger number of branches (see Fig.~\ref{fig:traindataset}). Consequently, there is a maximum $\nbr$, above which the addition of merger trees to the training dataset does not improve the reproduction of the progenitor mass. 
Therefore, with this new tool we would be able to predict with very good precision the most abundant merger trees, i.e. those of low- and intermediate-mass haloes at $z=0$ ($10^8\Msun< \Mvir <5\times10^{10} \Msun$), at least for the E100 simulation. 
Later on,  the  properties of the corresponding galaxies with $5\times10^5\Msun <  \mathrm{M_{galaxy}} <10^{8}\Msun$ could be obtained  (see e.g. \citealt{Gomez2021}),  using a SAM along with these merger trees. 

\begin{figure*}
    \centering
    \includegraphics[width=0.77\linewidth]{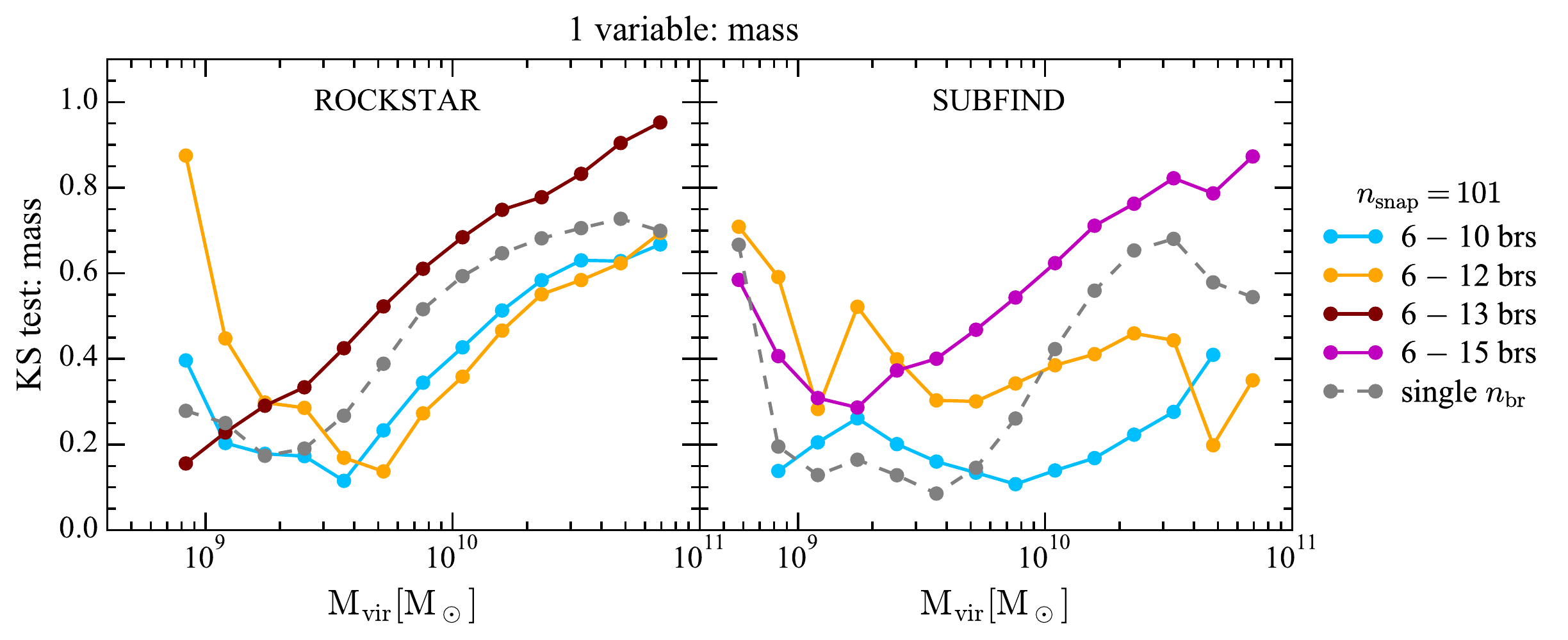}
    \includegraphics[width=0.77\linewidth]{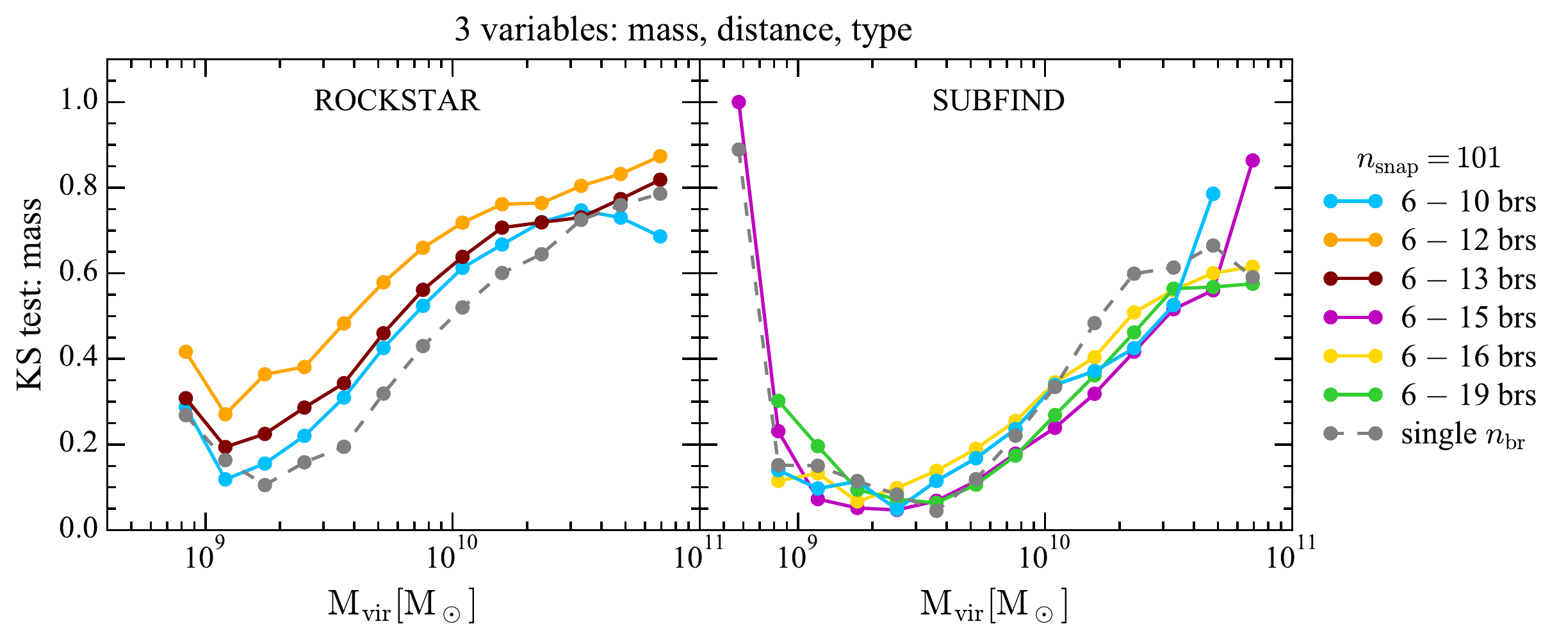}        
    \caption{
    Kolmogorov-Smirnov (KS) test for the mass of the progenitors and the final descendant for merger trees generated with 1 (top panels),  and 3 (bottom panels) input variables, 
    GAN model trained with either ROCKSTAR (left-hand panels) or SUBFIND (right-hand panels) merger trees. CDFs were constructed  for bins of the mass of the final descendant.  
    Grey lines correspond to series of individual trainings (single $\nbr$, one training per $\nbr$) and colourful lines to single trainings performed with  trees whose number of branches vary in a range (multiple $\nbr$, a single training per line). 
    }
    \label{fig:KStesthalomass}
\end{figure*}

Training our GAN model with different  combinations of variables, as well as with a single variable, is important to identify the most influential variables to predict the mass growth history of haloes.
In the second row of Fig.~\ref{fig:KStestmass}, we show similar trainings to those in the top panels, but performed with 2 variables: progenitor mass  and distance to the main branch, for $\nsnaps=101$. The test KS is performed only for the mass gain and loss cumulative distributions, i.e., we have evaluated if the addition of a second input, in this case the distance, helps to  more accurately reproduce the progenitor mass distribution. In the third row, we introduce the progenitor type instead of the distance to the main branch as second input. By comparing the second and third row with the top panels, we see that the introduction of a second variable improves SUBFIND results for the progenitor mass, but not those of ROCKSTAR, which in fact are worse than those obtained with the progenitor mass as the only input. As mentioned in the previous section, the addition of another input implies that the GAN model needs to learn to reproduce that input too, which affects the convergence of the training process. 
Regarding the SUBFIND learned representation, the best results for the KS test of the mass distribution are those where the progenitor type is the additional input, this is particularly true for trees with $6\leq\nbr\leq12$ (orange lines). For $6\leq\nbr\leq15$ (magenta line) the trend of the KS test is to increase with the number of branches when the type is the second variable and to decrease when the distance is considered instead. 
Note here that the progenitor type, unlike other inputs, is a discrete variable in our GAN model, so that in principle it should be easier to learn than the distance. 
Once again, in both cases ROCKSTAR and SUBFIND and for both distance and type, better results are obtained when a set of merger trees with number of branches in a given range is used as training data (solid lines) instead of individual trainings with fixed $\nbr$ (dashed grey lines). 

In the bottom panels of Fig.~\ref{fig:KStestmass}, we consider the three inputs at the same time in the training process. 
As expected from the results with two  inputs, ROCKSTAR learned representation of the progenitor mass is not improved by providing the GAN model with additional information about the progenitor type and distance to the progenitor in the main branch, with the sole exception of the results for $6\leq\nbr\leq13$ (brown lines).
For SUBFIND, we can see that the GAN model with three inputs, trained with halo merger trees with $6\leq\nbr\leq16$ (yellow line) yields the best progenitor mass distributions, in particular for ML trees with number of branches in the range $6\leq\nbr\leq11$, since there are more examples of these trees in the training dataset. These results are even better than those obtained with the progenitor mass as the only input and with mass and distance. Contrary to these cases (see panels above) the KS test tendency is to decrease with $\nbr$ in the aforementioned range, which in general holds for all the trainings with multiple $\nbr$. 
The second best results are those obtained with $6\leq\nbr\leq19$ (green line), for which the KS test increases for $\nbr\geq15$. 
Note that to achieve such a number of branches, we have tuned the size of the column- and row-wise filters, the input of the decoder and the batch size (see Table~\ref{tab:GANarch}). 
This is the maximum number of branches for which the GAN total loss function converges. Note however that the restriction on $\nbr^\mathrm{max}=19$ is imposed not only by the amount of trees in the training dataset, but mainly by memory constraints. For ROCKSTAR, on the other hand, $\nbr=13$ is the maximum number of branches that in this case is due to the amount of trees available for training, as previously mentioned.

Finally, in Fig.~\ref{fig:KStesthalomass} we show a global measure of the goodness of the reproduction of the mass gain-loss distribution of the progenitors across branches  by ML merger trees of haloes of a given virial mass, regardless of the specific number of branches. 
To that end, we construct CDFs of the above mentioned quantity as in Fig.~\ref{fig:metric_full_mass}, but instead of comparing samples of  merger trees with a given $\nbr$, we select `real' and ML trees by the mass of the final descendant.  
In Fig.~\ref{fig:KStesthalomass}, we depict the value of the test KS at the centre of each $\Mvir$ bin. From this figure, we immediately notice that as in Fig.~\ref{fig:KStestmass}, the best results are obtained for trainings carried out with 3 variables (bottom-right panel) for SUBFIND ML trees, and that for main haloes with masses in the $10^9 \Msun \lesssim \Mvir \lesssim 10^{11} \Msun$, the $6\leq\nbr\leq15$ (magenta line) multiple $\nbr$ training yields the best fit to the real trees. 
Note that for the first and last $\Mvir$ bins, the value of the KS test increases; this is due to the fact there are less merger trees of main haloes with masses in those ranges in the training dataset (see middle panel of Fig.~\ref{fig:traindataset}). 
For ROCKSTAR ML trees, on the other hand, we observe an improvement  on the KS test for merger trees generated with 3 variables (bottom-left panel) with respect to those trained with the mass as the sole input (top-left panel), for single $\nbr$ training. This was not evident in our results in Fig.~\ref{fig:KStestmass}. Moreover, when the GAN model receives 3 input variables the best fit to the `real'  mass gain-loss distribution is obtained with series of individual trainings per number of branches in the training dataset from $\nbr=6$ to $13$ (dashed grey line). Although, better results can be obtained with only one input variable for $\Mvir \gtrsim 3\times 10^{9}\Msun$; see orange and light blue lines in the top left panel of Fig.~\ref{fig:KStesthalomass}.

\subsection{Progenitor type}

\begin{figure*}
    \centering
    \includegraphics[width=0.9\linewidth]{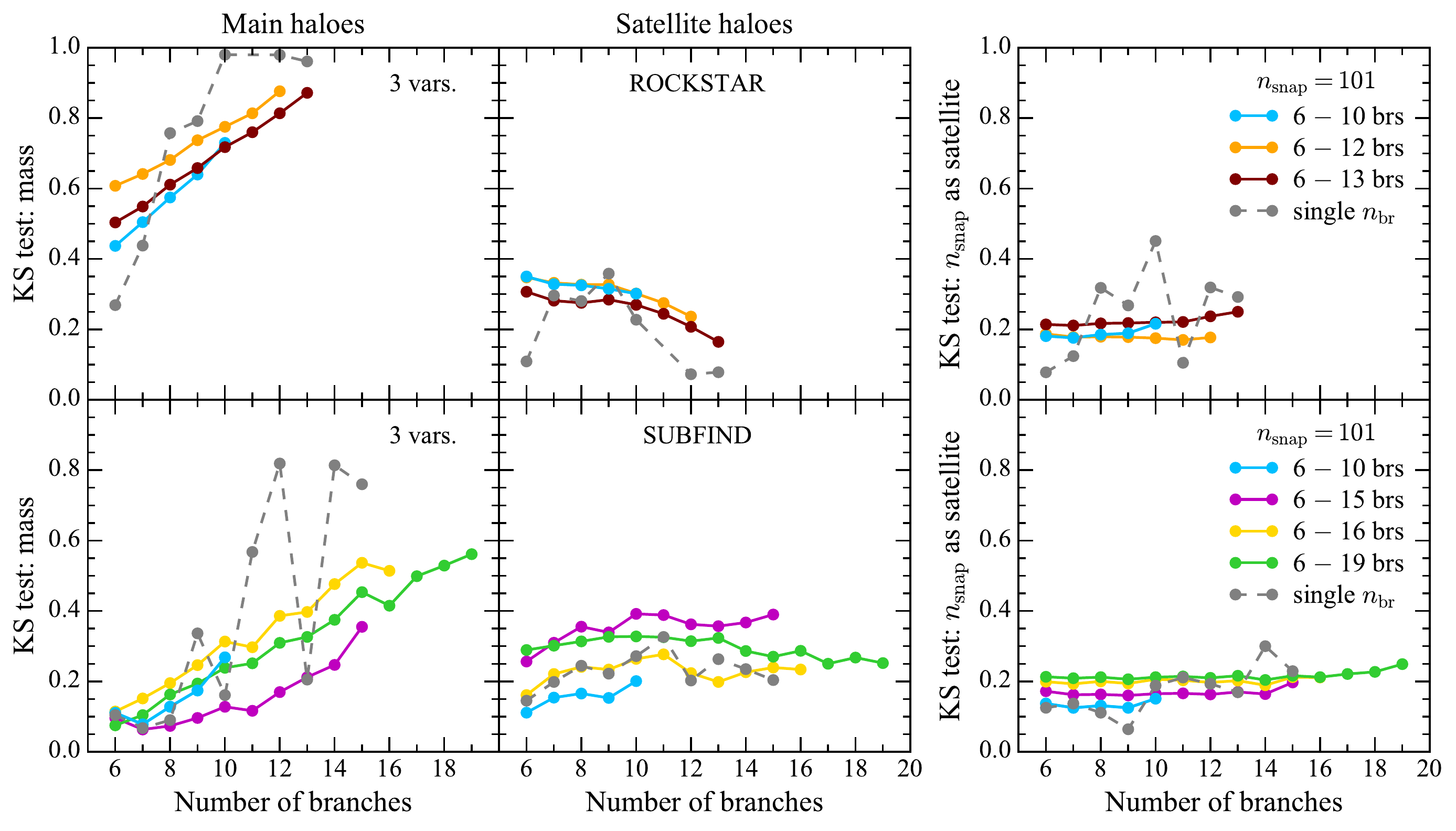}
    \caption{Kolmogorov-Smirnov (KS) test  for the mass of the progenitors per type: main haloes (left-hand panels) and satellites (middle panels) and for the number of snapshots a progenitor spend as a satellite (right-hand panels),  for merger trees generated with  3  variables, trainings performed with either ROCKSTAR (top panels) or SUBFIND (bottom panels) merger trees. Grey lines represent series of individual trainings (single $\nbr$, one training per $\nbr$) and colourful lines single trainings performed with  trees whose number of branches vary in a range (multiple $\nbr$, a single training per line). }
    \label{fig:KStesttype}
\end{figure*}

Next, we analyse the importance of discriminating main from satellite haloes when comparing the progenitor mass gain and loss distributions. As we have seen, this is particularly relevant for SUBFIND\@.  In Fig.~\ref{fig:KStesttype}, we show the values of the Kolmogorov-Smirnov test for main (left-hand panels) and satellite (middle panels) haloes for series of  trainings with  3 variables. 
For both ROCKSTAR and SUBFIND, the KS test for the satellite mass distribution remains approximately flat, especially for 
$6\leq\nbr\leq19$ (green line). This is true even for individual trainings (grey lines). Conversely, for main haloes the KS test increases with $\nbr$. 
It is worth noting that most of the progenitors in a merger tree are expected to be main haloes. Hence a better reconstruction of the mass gain and loss distributions of this fraction of the progenitors will improve the KS test of the full sample, shown in Fig.~\ref{fig:KStestmass}. 
This is particularly evident in the case of the trainings with 2 variables (mass and type), not shown in Fig.~\ref{fig:KStesttype}, and in the case of ROCKSTAR ML merger trees; compare orange, light blue and brown  lines in the first and second rows of Fig.~\ref{fig:KStestmass} with their respective counterparts in the third row of the same figure.

In addition, we observe that with 3 inputs, the reconstruction of the progenitor mass for  main haloes identified by ROCKSTAR is improved for the $6\leq\nbr\leq13$ training (brown line) with respect to that for $6\leq\nbr\leq12$ (orange line) that yielded the best results with 1 and 2 variables when considering the full progenitor sample (see Fig.~\ref{fig:KStestmass}, left-hand panels in the first and third rows). For satellite haloes, the $6\leq\nbr\leq13$ training also minimises the KS test for both 2 and 3 variables. 
For SUBFIND, on the other hand, the best fit for main haloes is obtained with the  $6\leq\nbr\leq15$ dataset (magenta line), while for satellites the corresponding best fit is achieved with $6\leq\nbr\leq10$ (light blue). Therefore, comparing the main halo result  with the bottom right panel of Fig.~\ref{fig:KStestmass} where better fits are obtained with the  $6\leq\nbr\leq16$ dataset (yellow line), we can infer that the inclusion of the distance between merging progenitors plays a greater role in SUBFIND learned representation of the progenitor mass, when the GAN is trained using datasets that contain 3 variables, than in that of ROCKSTAR.

Adding input channels to our GAN model implies they are also outputs of the neural network, as such they contribute to the GAN total loss and affect the convergence of the training process. Consequently, their fair reproduction validated against the training dataset  should also be assessed. 
In the right-hand panels of Fig.~\ref{fig:KStesttype}, we evaluate the fair reproduction of the progenitor type with 3 variables, using CDFs of the number of snapshots a progenitor is a satellite, see e.g., Fig.~\ref{fig:N_sns_as_sh}. It is worth noting that when considering only mass and progenitor type as inputs, the best results are obtained, in general,  when performing individual trainings with merger tree samples with a fixed number of branches, for both ROCKSTAR and SUBFIND\@. Adding the distance to the main branch as an extra input, improves the KS test for  trainings with multiple $\nbr$ and the opposite occurs for single $\nbr$ trainings in most cases. For SUBFIND, the number of snapshots that a progenitor is a satellite is  better reproduced by the training performed with the $6\leq\nbr\leq10$  dataset (light blue line), followed by the results for $6\leq\nbr\leq15$ (magenta line).  The KS test of the latter is very similar to the best ROCKSTAR result, obtained using merger trees with $6\leq\nbr\leq12$ (orange line). 

Despite our results in Fig.~\ref{fig:KStesttype}  favour the SUBFIND learned representation, it is worth remarking that progenitors identified by SUBFIND suffer from main halo-satellite switching issues \citep{Behroozi2015,Poole:2017}. As a merging process evolves and the progenitors involved approach each other, for instance a main halo-satellite pair (or more progenitors) identified as such at a given snapshot, at the next step in time they can be misidentified. This is the main halo is identified as a satellite, while the satellite becomes the main halo.  This can occur several times as the progenitors move closer. Since we do not correct for this issue, it is transferred to the learned representation. 

\subsection{Distance to the main branch}

\begin{figure*}
    \centering
    \includegraphics[width=0.77\linewidth]{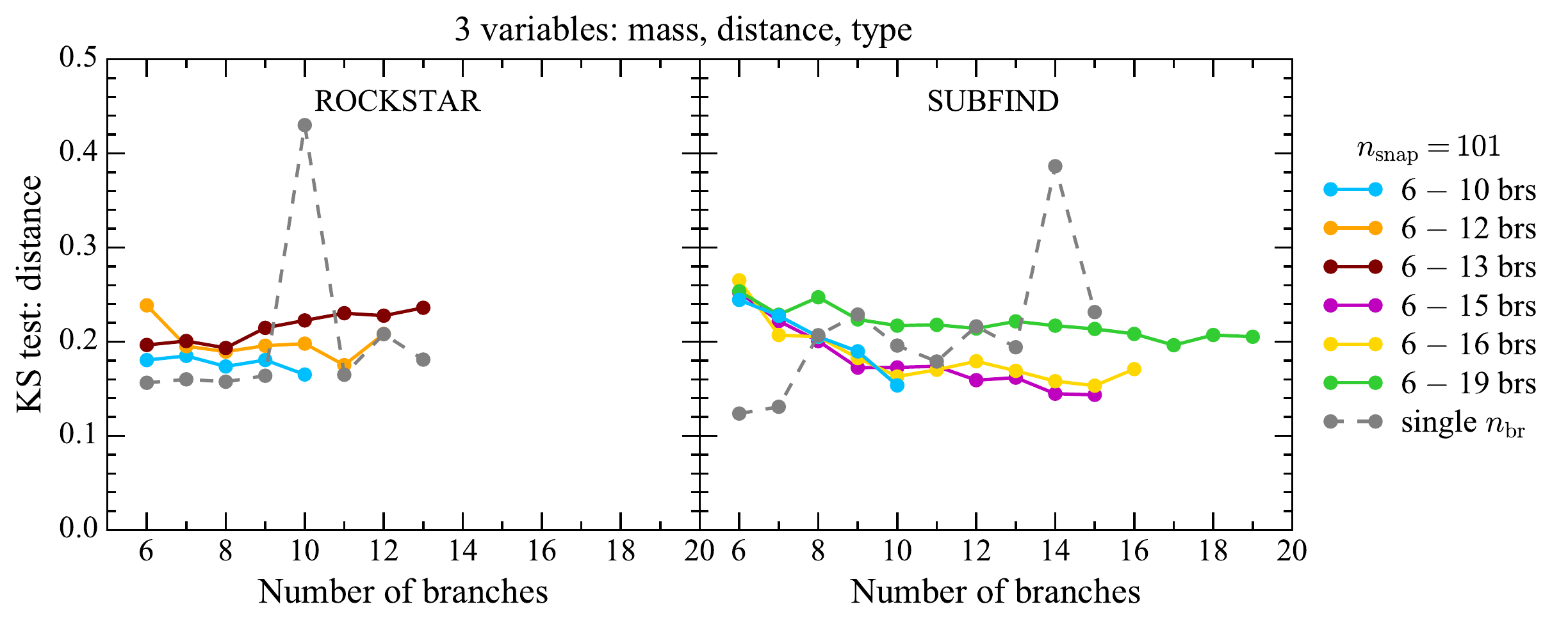}        
    \caption{Snapshot averaged Kolmogorov-Smirnov (KS) test for the distance of the progenitors to that in the main branch at the snapshot before the fusion occurs,  for merger trees generated with 3 variables,  trainings performed with either ROCKSTAR (left-hand panel) or SUBFIND (right-hand panel) merger trees. }
    \label{fig:KStestdistance}
\end{figure*}

Finally, we evaluate the fair reproduction of the distance of a progenitor to that in the main branch. In this case, we compare CDFs of this distance at the snapshot before the merger takes place. This snapshot varies from $0$ to the last snapshot, so that there is a probability distribution per snapshot, see e.g. Fig.~\ref{fig:distancedistrib}. Each distribution  is normalised before constructing the corresponding CDF. As a result, we obtain as many values of the KS test as snapshots in the  matrix representation of the merger trees. 
In Fig.~\ref{fig:KStestdistance}, we show these values averaged over this number of snapshots for series of trainings with 3 inputs. 
The best fits when comparing  averaged values are found for  trainings with multiple $\nbr$, specifically  $6\leq\nbr\leq10$ for ROCKSTAR (light blue line) and $6\leq\nbr\leq15$ for SUBFIND (magenta line), which unlike the case of the mass of the progenitors are very similar to the results obtained with single trainings with a fixed number of branches. 
Note however that when including the standard deviation of the KS test in Fig.~\ref{fig:KStestdistance}, there is in general good agreement among all trainings, especially for the 3 variables case. 
When considering only mass and distance as input variables, there is a more noticeable difference in the value of the KS test among different trainings. 
In addition, we find that the distance to the main branch is better reproduced when the GAN model is trained with merger trees constructed using ROCKSTAR. The difference being more striking for trainings with 2 variables.

\subsection{Length of the main branch}

\begin{figure}
\centering
\includegraphics[width=0.85\linewidth]{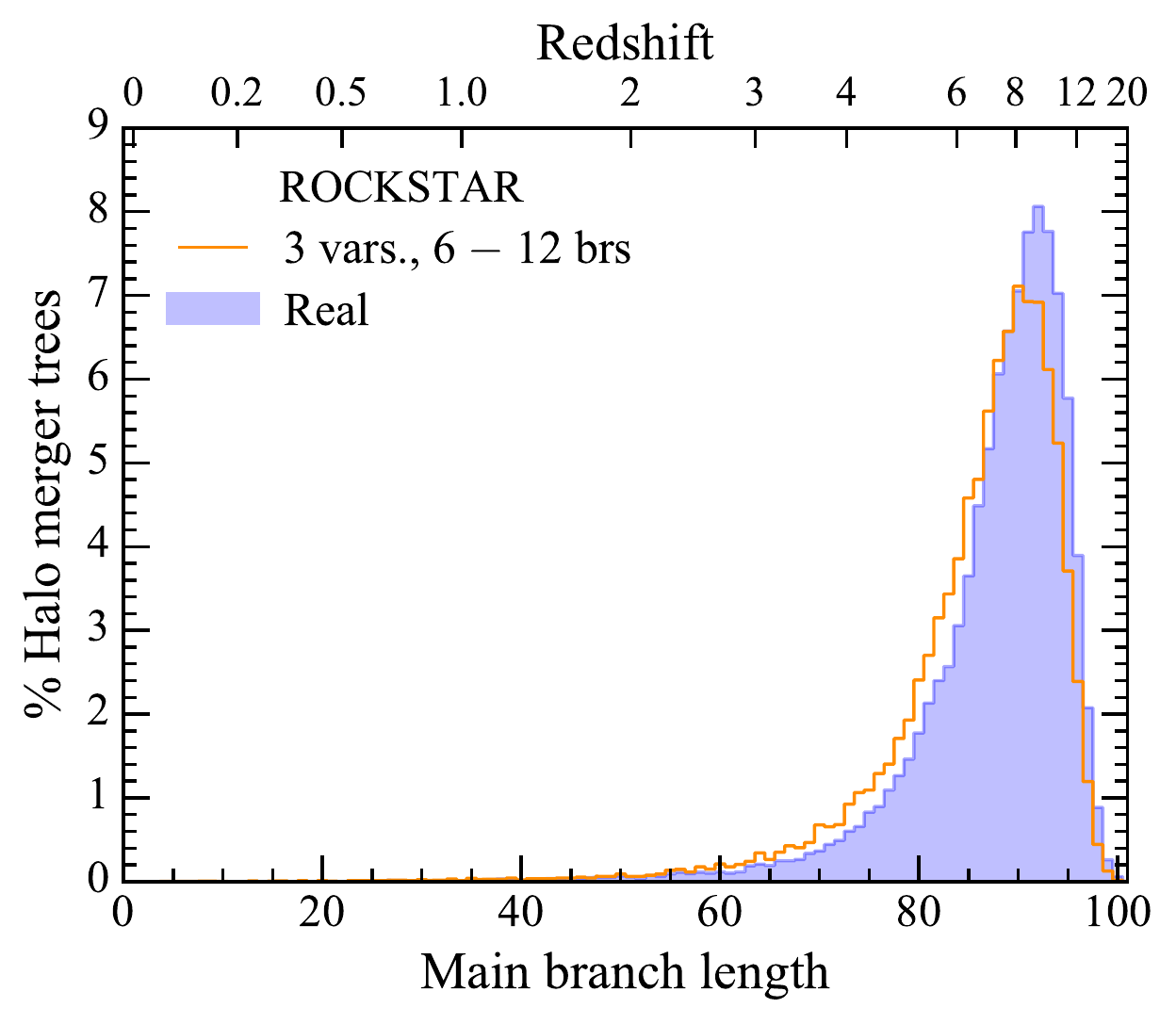}
\includegraphics[width=0.85\linewidth]{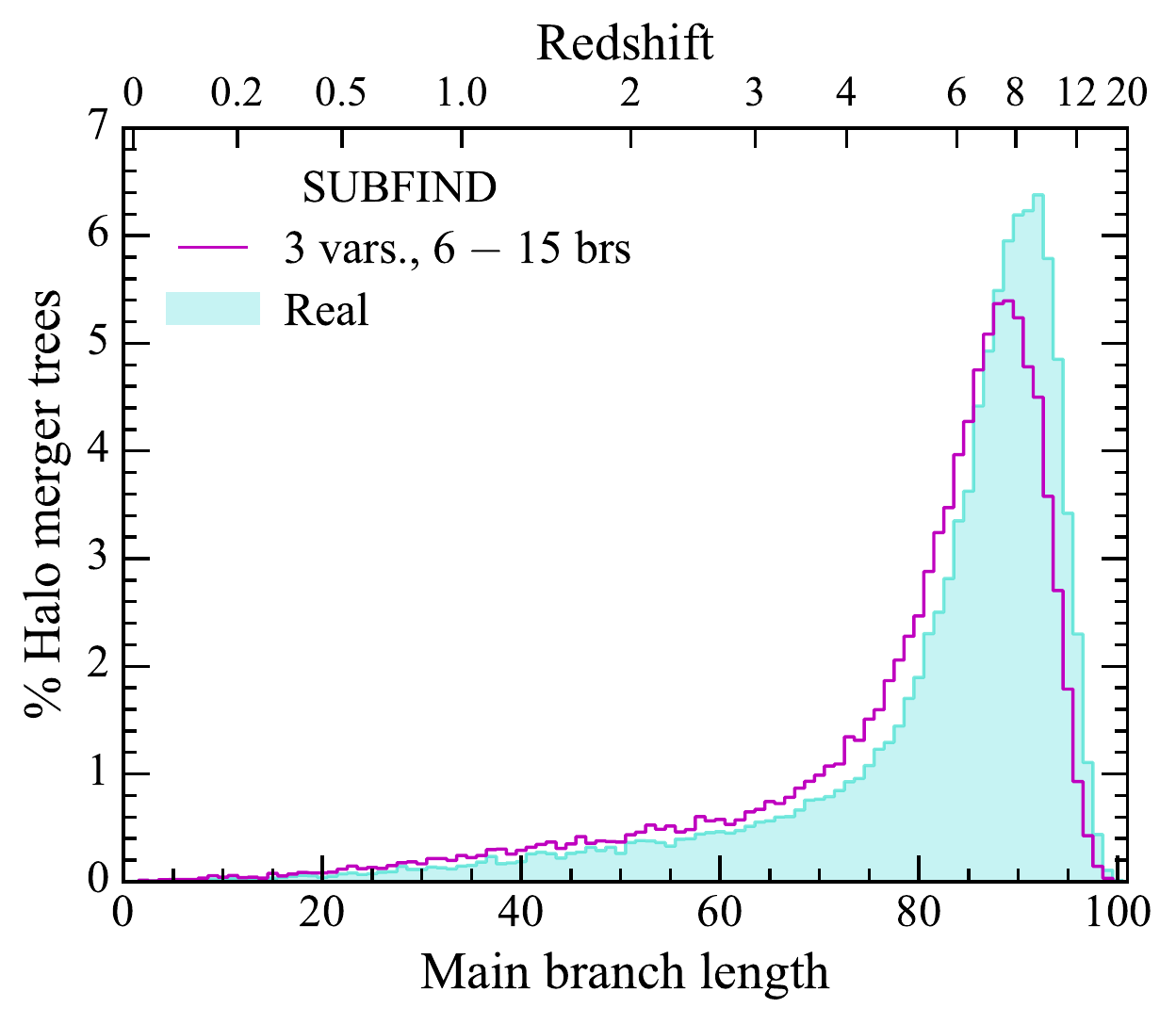}
\caption{
Histogram of the length of the main branch  measured in number of snapshots for machine learning (ML) merger trees, generated with 3 variables, a temporal resolution of  $\nsnaps=101$ and using the ROCKSTAR (top panel) training dataset that contains trees with $6-12$ branches (orange), and the dataset of  SUBFIND (bottom panel) trees with $6-15$ branches (magenta). For comparison, we also show the corresponding  histograms for samples of similar size of  `real' trees with the same number of branches. The redshift axis is also shown for reference. 
} 
  \label{fig:tmainbrlengthML}
\end{figure}

As we have shown in section~\ref{sec:mass}, trainings with 101 snapshots render a more accurate picture of the mass growth of haloes, when comparing ML merger trees with those in the training dataset. However,  this shortens the length of the main branch, which, as we shall see,  is not necessarily a shortcoming. 
In Fig.~\ref{fig:tmainbrlengthML}, we show a histogram of the main branch length for ML trees generated in multiple $\nbr$ trainings, with $\nsnaps=101$ and three input variables.   
We only show  ML trees generated during two trainings, specifically those carried out with the 
$6\leq \nbr \leq 12$ dataset for ROCKSTAR (top panel) and the $6\leq \nbr \leq 15$  dataset for SUBFIND (bottom panel), 
since  all multiple $\nbr$ trainings with 3 variables yield similar results. 
These are  the samples of ML trees used to construct probability distribution functions for the KS tests in Figs.~\ref{fig:KStestmass}--\ref{fig:KStestdistance}. We also show the corresponding histograms for the samples of  merger trees from the training dataset that were used to construct the CDFs of the `real' trees.
We can see that despite having reduced the number of snapshots,  our GAN model is still able to produce merger trees that trace progenitors back  in time (high in redshift) as further as the trees in the training dataset (compare the peaks of the real with the ML trees, redshift axis).  The peak of both distributions is still close to  $z\simeq8$. 
This is due to the fact that, when reducing the temporal resolution of the merger trees in the training dataset from $\nsnaps=201$ to $101$, we skipped intermediate snapshots along   the complete halo growth history.



\section{Conclusions}
\label{sec:conclusions}

Halo merger trees encode the mass assembly history of haloes. A complete set of merger trees is a necessary ingredient for semi-analytic models (SAMs) of galaxy formation and evolution.  These models are best suited to compare theory with observations, especially with data from forthcoming surveys, since they are computationally inexpensive when compared to cosmological hydrodynamical simulations. 
Currently,  the most common method to construct merger trees is based on computationally intensive N-body simulations. 
In this paper, we have shown that Generative Adversarial Networks (GANs), a deep learning framework, can be used to successfully learn and build  halo merger tree structures from cosmological N-body simulations. Their only limitations being memory resources and a sufficiently large training dataset.  The former can be circumvented by reducing the temporal resolution, while the latter depends on the the mass of the final descendant and the number of branches of the tree, features that ultimately depend on the volume of the dark matter only (DMO) simulation and the halo finder--tree builder algorithm. 
The main advantage of the machine learning (ML) generated trees is that they are  produced with a modest computational expense in a short computation time, while preserving the best features of merger trees from cosmological N-body simulations. This opens up the possibility to use ML trees, along with SAMs, to simulate large samples of galaxies comparable to those to be obtained by upcoming surveys, in a comparatively much shorter computational time than that required by cosmological hydrodynamical simulations.

We trained our GAN model with merger trees from the EAGLE simulation suite, constructed using two halo finder--tree builder algorithms:  SUBFIND--D-TREES and ROCKSTAR--{\sc ConsistentTrees}. 
We conducted a series of experiments designed to test the capabilities of our neural network model and to study the importance of including other variables in the training process aside from the mass of the progenitors and the final descendant. These additional input variables are the progenitor type, i.e. the condition of being a main halo or a satellite, and the distance of a merging progenitor to that in the main branch. Remarkably, the same GAN architecture with few changes in its parameters can be used to learn to generate merger trees obtained with both algorithms and with and without keeping fixed the number of branches. 

To evaluate the fair reproduction of the main properties of halo  merger trees,  we have compared probability distributions of equally large samples of ML-generated trees with those from the EAGLE DMO simulation with co-moving cubic box length of 100 Mpc, `real' trees, using the Kolmogorov-Smirnov test. The examined features include mass gain and loss of progenitors along a branch, number of snapshots a progenitor is a satellite, and distance between two merging progenitors at the snapshot before the merger event, and the interplay of the above. Note that more properties of the merging progenitors can also be learned just by adding the corresponding quantities as additional input channels and tuning the parameters of our GAN model.

In broad terms, our GAN model learns to generate halo merger trees identified by both pairs of halo finder--tree builder algorithms. When evaluating the quality of both learned representations, 
we find that our GAN model is more successful in generating SUBFIND-like merger tree structures than in accurately reproducing the statistical features of the bulk of `real' ROCKSTAR merger trees. This is clearly influenced by the fact that there are more SUBFIND merger trees available for training per given number of branches in our training database. 
We also find that including the progenitor type as an input of the training process helps our GAN model to predict with more precision the mass gain and loss of the progenitors that are main haloes, especially for SUBFIND-like ML merger trees. Note that this is particularly relevant for ML trees intended to be used with SAMs of galaxy formation, like GALFORM, SAGE, SAG and GALACTICUS, as both  mass growth and progenitor type are arguably the two most important inputs for galaxy evolution and to derive galactic properties such as the hot and cold gas mass fractions and the star formation rate.

Adding the distance of a progenitor to that in the main branch also improves SUBFIND learned representation of the halo mass assembly history. It is worth noting than when introducing an  extra input, our GAN model must also learn to reproduce this new variable. In this regard, when evaluating the well reproduction of the statistical distributions of the distance of a progenitor to that in the main branch, we find that  ROCKSTAR learned representation follows more accurately the distributions of the `real' sample. Although, this distance is not an essential ingredient of SAMs, some of them like GALFORM  take these distances into account  to redefine the progenitor type of an infalling halo; hence, being able to predict this variable is crucial for these SAMs.

Finally, our GAN-based halo merger tree generation framework can be used to construct merger histories of haloes of low and intermediate mass, for which there is, in general, a large number of samples available in cosmological simulations. Massive haloes, on the other hand, have a more intricate assembly history and appear less frequently. As mentioned above, ML merger trees can be used in SAMs to model galaxy formation and  evolution, keeping in mind the lack of rare populations corresponding to massive haloes, which can be alleviated by training our  GAN model   using merger trees from simulations of larger volume than that of the EAGLE simulation suite.
According to the specific requirements of a given SAM, more  input variables such as position, peculiar velocity and velocity dispersion of the progenitors, among others, can be easily added and learned by our GAN model.

\section*{Acknowledgements}

SR was partially supported by MINECO/FEDER (Spain) under grant   PGC2018-094975-C2,  by the UK STFC grant ST/T000759/1, and by the Australian Research Council through the ARC Centre of Excellence for Dark Matter Particle Physics, CE200100008. 
SR also acknowledges funding from the European Union's Horizon 2020 Research and Innovation Programme
under the Marie Sklodowska-Curie grant agreement No. 734374 (LACEGAL-RISE) for a secondment at the  Pontificia Universidad Cat\'olica de Chile.
JSG acknowledges support from CONICYT project Basal AFB-170002, funding  from  the CONICYT PFCHA/DOCTORADO BECAS CHILE/2019 21191147, and the Predoctoral contract `Formación de Personal Investigador' from the Universidad
Autónoma de Madrid (FPI-UAM, 2021). 
ARR acknowledges support from the Brazilian National Council for Scientific and Technological Development (CNPq) under grant No.~307425/2017-7.  ARR was also with University of Campinas and University of Reykjavik while working on this research. 
This research was undertaken using the LIEF HPC-GPGPU Facility hosted at the University of Melbourne. This Facility was established with the assistance of LIEF Grant LE170100200. 
This work also used computer facilities at the Universidad Aut\'onoma de Madrid and the Geryon computer at the Center for Astro-Engineering UC, part of the BASAL PFB-06, which received additional funding from QUIMAL 130008 and Fondequip AIC-57 for upgrades.
The authors would also like to thank the CYTED AgIoT Project (520rt0011), CORFO CoTH2O Consortium, and Proyecto Asociativo UDP `Plataformas Digitales como modelo organizacional', for their support. 


\section*{Data Availability}

The data underlying this article will be shared on reasonable request to the corresponding author. Merger trees from the EAGLE simulation suite obtained with SUBFIND have been publicly released~\citep{McAlpine2016}.


\bibliographystyle{mnras}
\bibliography{references}



\appendix

\section{GAN architecture}
\label{sec:GANarchitecture}

\begin{table}
\caption{Layout and parameters of the discriminator, encoder and decoder networks, where $\nsnaps$ denotes the maximum number of snapshots (temporal resolution), $\nbr$ the maximum number of branches and $\nvar$ the number of variables, k is the kernel structure and s the number of strides.}
\label{tab:GANarch}
\centering
\begin{tabular}{|lcl|}
  \hline
  \multicolumn{3}{c}{Discriminator}\\
  \hline
  Layer & Parameters & Output shape \\ 
  \hline
  Input  &  & ($\nsnaps$,$\nbr$,$\nvar$) \\
  Conv2D & k:(1,5--9) \,s:1  &($\nsnaps$,$\nbr$,16) \\
  ELU &  &($\nsnaps$,$\nbr$,16) \\
  Conv2D & k:(1,5--9) \,s:1 & ($\nsnaps$,$\nbr$,32) \\
  ELU &  & ($\nsnaps$,$\nbr$,32) \\
  Conv2D & k:(3--5,1) \,s:1 & ($\nsnaps$,$\nbr$,64) \\
  ELU &  & ($\nsnaps$,$\nbr$,64) \\
  Conv2D & k:(3--5,1) \,s:1 & ($\nsnaps$,$\nbr$,128)\\
  ELU &  & ($\nsnaps$,$\nbr$,128) \\
  Conv2D & k:(3--5,1) \,s:1 & ($\nsnaps$,$\nbr$,256)\\
  ELU &  & ($\nsnaps$,$\nbr$,256) \\
  Flatten & & ($\nsnaps\times \nbr \times 256$) \\ 
  FC &  & (1) \\
  \hline
  \multicolumn{3}{c}{Encoder}\\  
  \hline
  Layer & Parameters & Output shape \\ 
  \hline
  Input  &  & ($\nsnaps$,$\nbr$,$\nvar$) \\
  Conv2D & k:(1,5--9) \,s:1  &($\nsnaps$,$\nbr$,16) \\
  ELU &  &($\nsnaps$,$\nbr$,16) \\
  Conv2D & k:(1,5--9) \,s:1 & ($\nsnaps$,$\nbr$,32) \\
  ELU &  & ($\nsnaps$,$\nbr$,32) \\
  Conv2D & k:(3--5,1) \,s:1 & ($\nsnaps$,$\nbr$,64) \\
  ELU &  & ($\nsnaps$,$\nbr$,64) \\
  Conv2D & k:(3--5,1) \,s:1 & ($\nsnaps$,$\nbr$,128)\\
  ELU &  & ($\nsnaps$,$\nbr$,128) \\
  Conv2D & k:(3--5,1) \,s:1 & ($\nsnaps$,$\nbr$,256)\\
  ELU &  & ($\nsnaps$,$\nbr$,256) \\
  Flatten & & ($\nsnaps\times \nbr \times 256$) \\ 
  FC &  & (100--300) \\
  \hline
  \multicolumn{3}{c}{Decoder}\\  
   \hline
  Layer & Parameters & Output shape \\ 
  \hline
  Input  &  & (100--300) \\
  FC &  & ($\nsnaps\times \nbr \times 256$)\\
  ELU &  & ($\nsnaps\times \nbr \times 256$) \\
  Deconv2D & k:(5--9,1) \,s:1  &($\nsnaps$,$\nbr$,128) \\
  ELU &  &($\nsnaps$,$\nbr$,128) \\
  Deconv2D & k:(5--9,1) \,s:1 & ($\nsnaps$,$\nbr$,64) \\
  ELU &  & ($\nsnaps$,$\nbr$,64) \\
  Deconv2D & k:(1,3--5) \,s:1 & ($\nsnaps$,$\nbr$,32) \\
  ELU &  & ($\nsnaps$,$\nbr$,32) \\
  Deconv2D & k:(1,3--5) \,s:1 & ($\nsnaps$,$\nbr$,16)\\
  ELU &  & ($\nsnaps$,$\nbr$,16) \\
  Deconv2D & k:(1,3--5) \,s:1 & ($\nsnaps$,$\nbr$,$\nvar$)\\
  Sigmoid &  & ($\nsnaps$,$\nbr$,$\nvar$) \\
  \hline
\end{tabular} 
\end{table}

As mentioned in section~\ref{sec:GANmodel}, the GAN architecture is based on the model introduced by \citet{Robles:2019nfk}, whose layout is implemented using a combination of CNNs layers with either column- or row-like filters. 
We keep the same activation function between layers, namely the Exponential Linear Unit (ELU), and calculate the losses (classic GAN loss and reconstruction losses per input channel) with cross entropy with logits and sigmoid activation. These choices produced the best results in \citet{Robles:2019nfk}. 
The parameters of the discriminator, encoder and decoder are given in Table~\ref{tab:GANarch}. 
Note that  we have indicated a range for the kernel (k) size of the filters. The specific size depends on the particular training dataset, namely the halo finder--tree builder algorithm used to construct them, ROCKSTAR or SUBFIND, the maximum number of branches and time resolution  considered. 
We have found that for our GAN model to successfully learn to construct SUBFIND-like merger trees, a larger kernel size for  all filters is always required. 
Both, the discriminator and generator, are trained with batches of merger trees, using the Adam optimiser.   
The optimal batch size varies in the range of 100--200 samples and  depends mainly on the maximum number of branches in the training dataset, subject to memory resources. Finally, the input size of the decoder (output of the encoder) also changes according to the number of columns in the matrix representation.

\section{Examples of generated merger trees}
\label{sec:examples}

\begin{figure*}
\centering
\includegraphics[width=\columnwidth]{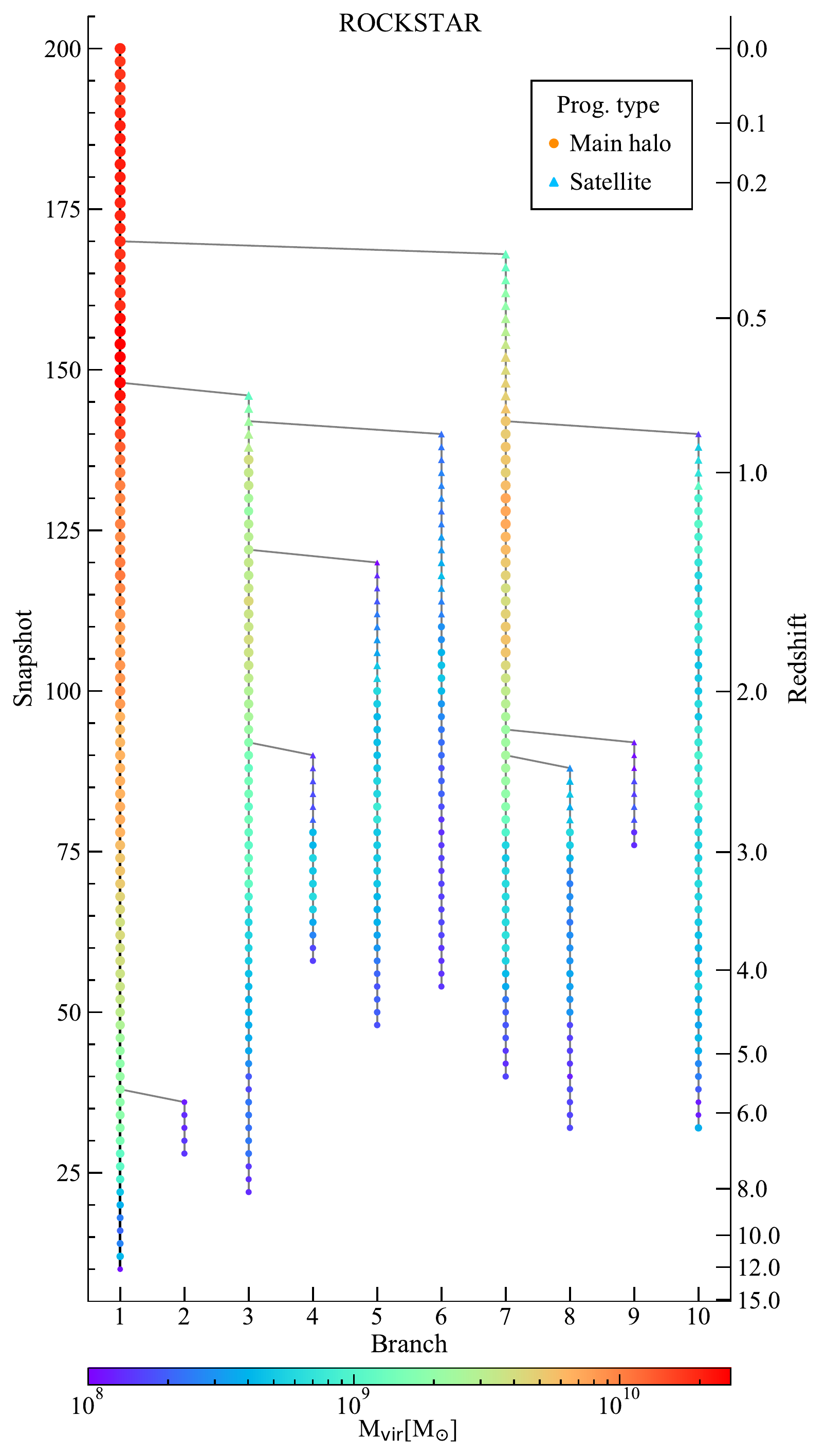}
\includegraphics[width=\columnwidth]{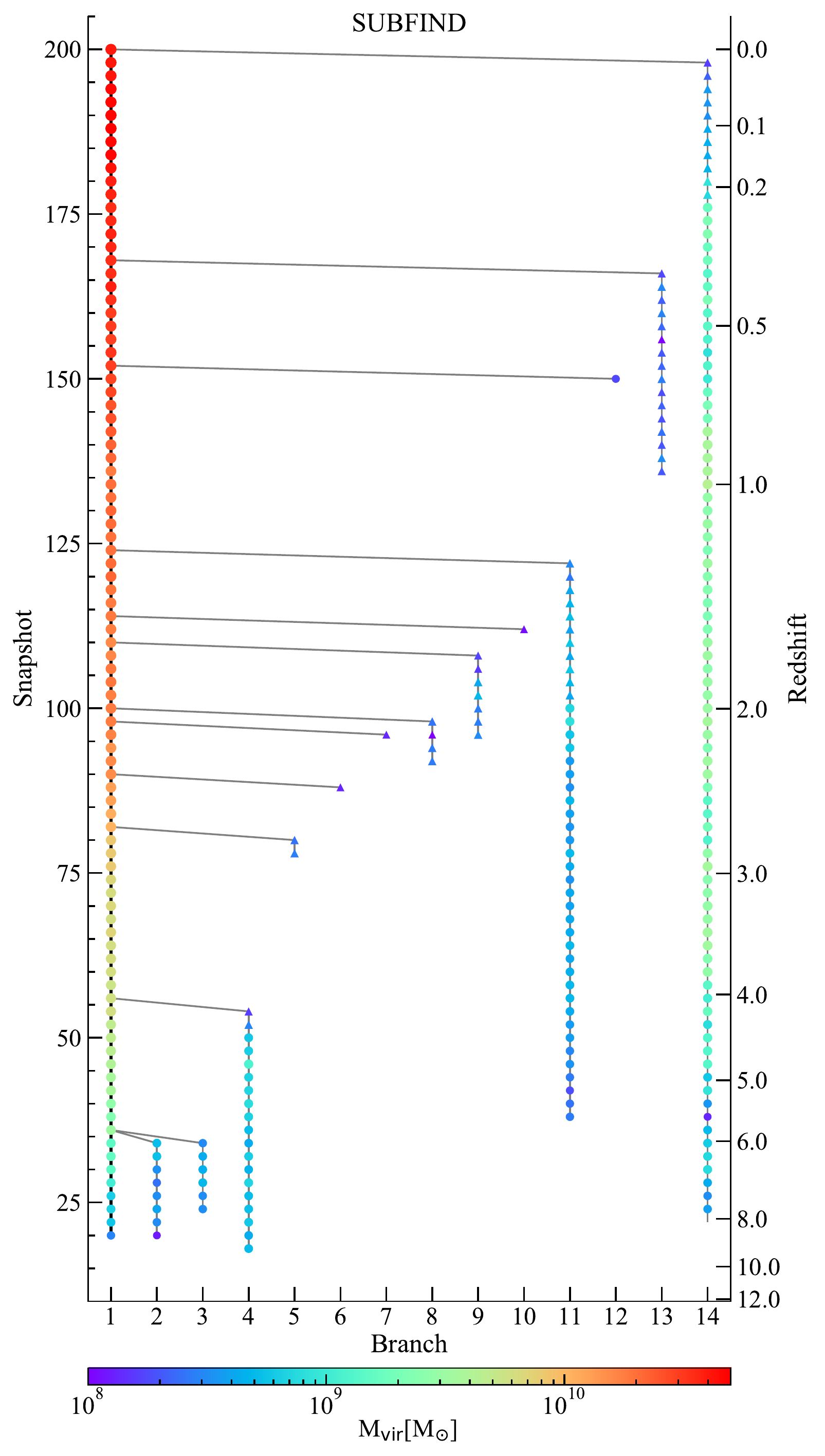}
\caption{
  Additional examples of ROCKSTAR-like (left-hand panel, 10 branches) and SUBFIND-like (right-hand panel, 14 branches) ML-generated merger trees, showing  the tree structure (snapshot vs.\ branch). 
  Progenitors that are main haloes are denoted by  circles and satellites by triangles,  the colour map represents the virial mass of the halo progenitors. 
 }
  \label{fig:mergertreestruct}
\end{figure*}

In Fig.~\ref{fig:mergertreestruct}, we show more examples of generated merger trees in the classic tree representation, snapshot vs.\ branch. 
These halo merger trees were obtained after training our GAN model with three variables in the same multiple $\nbr$ training as those in Fig.~\ref{fig:mergertree}, namely $6-10\,\nbr$ (ROCKSTAR, light blue lines in the left-bottom panels of Figs.~\ref{fig:KStestmass} -- \ref{fig:KStestdistance}) and $6-16\,\nbr$ (SUBFIND, yellow lines in the right-bottom panels of Figs.~\ref{fig:KStestmass} -- \ref{fig:KStestdistance}) and time resolution $\nsnaps=101$. Due to the more complex structure of these trees, we do not show the corresponding plots  in the plane snapshot vs.\ distance to the main branch. Note that in the ROCKSTAR-like merger tree (left), there are two branches which have subbranches, in contrast to that of SUBFIND-like tree (right). In the particular halo mass range we have focused due to the selection criterion for the training dataset, i.e. $10^{9}\Msun \lesssim \Mvir \lesssim 10^{11}\Msun$, we note that on average ROCKSTAR merger trees tend to have more subbranches than those identified with SUBFIND as well as longer branches (excluding the main branch).


\bsp	
\label{lastpage}
\end{document}